\documentclass[iop]{emulateapj}
\pdfoutput=1

\slugcomment{Accepted for publication in ApJ}

\newcommand{\um}{~\mu m}
\newcommand{\msun}{$~M_{\odot}$}
\newcommand{\dgr}{^\circ}

\usepackage{natbib}
\usepackage{url}

\shorttitle{Environments of USco Brown Dwarf Companions}
\shortauthors{Bailey et al.}

\begin{document}

\title{A Thermal Infrared Imaging Study of Very Low-Mass, Wide Separation Brown Dwarf Companions to Upper Scorpius Stars: Constraining Circumstellar Environments*}

\altaffiltext{*}{Observations reported here were obtained at the LBT and MMT Observatories. MMT Observatory is a joint facility of the University of Arizona and the Smithsonian Institution. The LBT is an international collaboration among institutions in the United States, Italy and Germany. LBT Corporation partners are: The University of Arizona on behalf of the Arizona university system; Istituto Nazionale di Astrofisica, Italy; LBT Beteiligungsgesellschaft, Germany, representing the Max-Planck Society, the Astrophysical Institute Potsdam, and Heidelberg University; The Ohio State University, and The Research Corporation, on behalf of The University of Notre Dame, University of Minnesota and University of Virginia.}

\author{Vanessa Bailey\altaffilmark{1}, 
Philip M.\ Hinz\altaffilmark{1},  
Thayne Currie\altaffilmark{2},
Kate Y.\ L.\ Su\altaffilmark{1}, 
Simone Esposito\altaffilmark{3}, 
John M.\ Hill\altaffilmark{4}, 
William F.\ Hoffmann\altaffilmark{1}, 
Terry Jones\altaffilmark{5}, 
Jihun Kim\altaffilmark{6}, 
Jarron Leisenring\altaffilmark{7}, 
Michael Meyer\altaffilmark{7},
Ruth Murray-Clay\altaffilmark{8}, 
Matthew J.\ Nelson\altaffilmark{9},
Enrico Pinna\altaffilmark{3}, 
Alfio Puglisi\altaffilmark{3}, 
George Rieke\altaffilmark{1}, 
Timothy Rodigas\altaffilmark{1}, 
Andrew Skemer\altaffilmark{1}, 
Michael F.\ Skrutskie\altaffilmark{8}, 
Vidhya Vaitheeswaran\altaffilmark{1}, 
and John C.\ Wilson\altaffilmark{9}
\email{vbailey@as.arizona.edu}}

\affil{$^1$ Steward Observatory, Department of Astronomy, University of Arizona}
\affil{$^2$ Department of Astronomy and Astrophysics, University of Toronto}
\affil{$^3$ Istituto Nazionale di Astrofisica, Osservatorio Astrofisico di Arcetri}
\affil{$^4$ Large Binocular Telescope Observatory, University of Arizona}
\affil{$^5$ School of Physics and Astronomy, University of Minnesota}
\affil{$^6$ College of Optical Sciences, University of Arizona}
\affil{$^7$ Institute for Astronomy, ETH Zurich}
\affil{$^8$ Harvard-Smithsonian Center for Astrophysics}
\affil{$^9$ Department of Astronomy, University of Virginia}

\begin{abstract}
We present a $3-5\um$ LBT/MMT adaptive optics imaging study of three Upper Scorpius stars with brown dwarf (BD) companions with very low-masses/mass ratios ($M_{BD}$ $<25~M_{Jup}$; $M_{BD}$/$M_{\star}$ $\approx$ 1--2\%),  and wide separations ($300-700~AU$): GSC~06214, 1RXS~1609, and HIP~78530. We combine these new thermal IR data with existing $1-4\um$ and $24\um$ photometry to constrain the  properties of the BDs and identify evidence for circumprimary/secondary disks in these unusual systems.  We confirm that GSC~06214B is surrounded by a disk, further showing this disk produces a broadband IR excess due to small dust near the dust sublimation radius. An unresolved $24\um$ excess in the system may be explained by the contribution from this disk.  1RXS~1609B exhibits no $3-4\um$ excess, nor does its primary; however, the system as a whole has a modest $24\um$ excess, which may come from warm dust around the primary and/or BD. Neither object in the HIP 78530 system exhibits near- to mid-IR excesses. We additionally find that the $1-4\um$ colors of HIP~78530B match a spectral type of M$3\pm2$, inconsistent with the M8 spectral type assigned based on its near-IR spectrum, indicating it may be a low-mass star rather than a BD. We present new upper limits on additional low-mass companions in the system ($<5~M_{Jup}$ beyond $175~AU$). Finally, we examine the utility of circumsecondary disks as probes of the formation histories of wide BD companions, finding that the presence of a disk may disfavor BD formation near the primary with subsequent outward scattering. \end{abstract}


\section{Introduction}

Nearly all stars are born with optically thick, gas-rich, dusty accretion disks that are thought to be analogous to the early solar nebula and comprise the building blocks of planet formation. Within $5-10~Myr$, most of these disks disappear, having presumably converted their mass into larger, planetesimal-to-planet-mass bodies. The timescale for this process typically decreases with increasing stellar mass \citep{Carpenter2006a,Currie2009}. The youngest very low-mass stars and brown dwarfs (BD) are also typically surrounded by disks. Their disk lifetimes also tend to increase with decreasing BD mass \citep{Luhman2008, Carpenter2009, Monin2010}, though some work suggests this behavior may not be universal \citep{Luhman2010} (hereafter L2010). 

While many young, single BDs, like stars, are surrounded by optically-thick accretion disks, it is as yet unclear whether all BDs formed like stars, in particular those that are companions to stars in wide binary systems. Between $10-30\%$ of BDs are believed to form in binary systems \citep{Burgasser2003, Close2003, Joergens2008}, a frequency significantly lower than that for low-mass M stars and solar-mass stars  ($\gtrsim$ 30\% and $\sim50\%$; \citealt{Fischer1992, Janson2012b, Raghavan2010}).   If wide BD companions formed like other, more massive stellar binary companions, they should comprise the tail end of the binary mass function (BMF). Previous studies of the stellar BMF have found that low mass ratios ($M_B/M_A < 10\%$) are rare \citep[e.g.][]{Reggiani2011}.   In contrast, recent observations suggest the frequency of at least solar-type stars with very low-mass BD companions may be higher than expected if such objects were drawn from an extrapolation of the stellar BMF \citep{Ireland2011, Janson2012a}.  Additionally, some of the lowest mass wide-separation companions have masses straddling or below the deuterium-burning limit often used to distinguish between planets and brown dwarfs \citep[e.g.][]{Luhman2006, Metchev2006}.   

In light of these complications, several other, more planet-like, formation mechanisms have been proposed for these wide, low-mass BD companions. In these scenarios the BDs form from the primary's protoplanetary disk, either in situ or initially much closer to the primary.  The first possibility is that the BDs form as part of a system of massive objects orbiting the primary and are subsequently scattered by one of these objects to their current location \citep[e.g.][]{Veras2009}. However, a dearth of close BD companions (the so-called ``brown dwarf desert'') has been shown observationally and suggested theoretically \citep{Marcy2000, Bate2000}. Another possibility is planet-like formation in situ via direct collapse within the primary star's protoplanetary disk \citep[e.g.][]{Boss2011}. However, the typical stellar protoplanetary disk radius is $20-200~AU$, so BDs formed at separations $>200~AU$ would require stellar hosts with unusually large disks \citep{Andrews2009}. Furthermore, the mass ratios and separations of wide BD companions are distinct from those of RV-detected planets as well as the directly-imaged planets/candidates around HR 8799, $\beta$ Pic, Fomalhaut, and $\kappa$ And \citep{Marois2008, Marois2010, Lagrange2010, Kalas2008, Currie2012, Carson2013},  indicating that they may comprise a distinct population with a separate formation mechanism \citep[e.g.][]{Kratter2010}. 

High-contrast imaging observations of these BD companions and constraints on their disk population may clarify their formation mechanism.  If wide BD companions at hundreds of AU separation form in situ as the tail end of the binary mass function (just like stellar companions), we expect that they have the same disk fraction as single BDs, since binarity at these wide separations appears to not affect circumstellar disk evolution \citep{Jensen1996, Pascucci2008, Kraus2012}. If they have a more planet-like formation mechanism, they are still likely to be born with disks. In our own Solar System, this is evidenced by the system of coplanar moons around Jupiter \citep{Lunine1982, canup2002}. However, if these BDs form initially much closer to the primary and are scattered to their current locations, their disk masses may not be great enough to survive to the present day, as we will demonstrate. Furthermore, an intermediate-mass perturber should be present close to the primary star and may be detectable in imaging data.

The nearby  Upper Scorpius OB Association (USco) provides a good testbed for these scenarios.  Its youth ($5-10~Myr$, \citealt{Preibisch2002, Pecaut2012, Song2012}) and large number of members ($N$ $\gtrsim$ 500, \citealt{Preibisch2002}) mean that a  significant population of massive disks remains, allowing for a robust statistical sampling of the association's disk population  from early to late-type stars and BDs \citep[][]{Carpenter2006a, Carpenter2009, Luhman2012}.  Its proximity ($145~pc$) provides good  sensitivity at low disk and companion masses \citep{DeZeeuw1999, Preibisch2002}.  About 25\% of the M-type BDs in USco exhibit broadband mid-infrared excess emission from warm dust in circum(sub)stellar disks \citep{Scholz2007, Riaz2009, Riaz2012, Luhman2012, Dawson2013}. Sensitivity limits have thus far precluded similar studies of less massive L-type members.
 
Furthermore, USco includes three stars with recently discovered wide-separation, very low mass (ratio) BD companions: GSC~06214-00210 (GSC~06214), 1RXS~160929.1-210524 (1RXS~1609), and HIP~78530; \citep{Ireland2011, Lafreniere2008, Lafreniere2011}, one of which, GSC~06214B, has been discovered to retain a hot circum(sub)stellar disk \citep{Bowler2011}. The first two systems both are K-type primaries and L-type BDs, while the third is a B-type primary and M-type BD. These three systems constitute the lowest known mass ratio binary systems in USco: $q$ $\approx$ 1--2\%, which is exceedingly rare in stellar binaries, regardless of the primary star's mass \citep[e.g.][]{Reggiani2011}. In fact these are among the highest mass contrast non-planetary systems found to date, populating the region of parameter space between planetary systems and stellar binaries. By probing the circum(sub)stellar environments of these and other similar systems, it may be possible to  discriminate between the various suggested formation histories; here we take the first steps towards that goal. 

To explore the circumstellar properties of these very low mass ratio pairs,  we present and analyze new adaptive optics-resolved near/thermal IR imaging in four bands [$3.1\um$, $3.3\um$, $L^\prime$ and $M^\prime$] as well as archival unresolved Spitzer/MIPS-$24\um$ photometry.  In each system we search for the presence of a hot circum(sub)stellar disk component around the BD, identifiable by broadband infrared excess emission in our array of resolved $1-5\um$ data.  This emission is due to small dust, with temperatures above several hundred Kelvin, located in the innermost (hot) disk where accretion is likely to be active \citep{Tuthill2001}.  The MIPS-$24\um$ data probes the presence of warm ($T_{eff}$ $\sim$ 100 $K$) dust in the system, although the large beam size of these observations precludes us from determining whether warm dust emission originates from around the primary or the secondary (or both).

Our study is organized as follows. Section \ref{sect:Observations} describes our sample in more detail, our observations, and our data reduction.  In Section \ref{sect:ResultsAnalysis}, we re-examine the companions' spectral types and assess their near and mid-IR excesses using these new data.  With our high-resolution adaptive optics (AO) imaging we are also able to provide constraints on the presence of additional companions in the systems. We further examine the disk properties of GSC~06214B and the nature of HIP~78530B in Section \ref{sect:Discussion}, discuss the utility of disks as probes of scattering histories in Section \ref{sect:DiskHistories}, and summarize our findings in Section \ref{Summary}.

%%%%%%%%%%%%%%%%%%%%%%%%%%%%%%%%%%%%%%%%%%%%%%%%%%%%%%%%%%%%%%%%%%

\section{Sample, Observations and Data Reduction}
\label{sect:Observations}
\subsection{Sample}

Here we summarize the basic system properties of the three newly-identified very low-mass (ratio) BD companions that are the subject of our study:

\begin{itemize}
\item GSC~06214A is a $0.9~M_{\odot}$ K7 star. Its membership is based on proper motion as well as on the youth indicators: X-ray emission, H$\alpha$ emission, and lithium absorption \citep[and references therein]{Bowler2011}. GSC~06214B orbits at $2.20''$ ($320~AU$) projected separation. \citet{Ireland2011} confirmed its common proper motion over a $3~yr$ baseline and assigned it a spectral type of L0 and a mass of $14\pm2~M_{Jup}$ based on $5~Myr$ DUSTY models \citep{Chabrier2000}. \citet{Bowler2011} spectroscopically confirmed the object, and further suggested it is still actively accreting from a protoplanetary disk based on a $4\um$ IR excess and Pa$\beta$ emission. 

\item 1RXS~1609A is another $0.9~M_{\odot}$ K7 star, with membership determined similarly to GSC~06214A \citep[and references therein]{Lafreniere2008}. 1RXS~1609B orbits at $2.22''$ ($320~AU$) projected separation. \citet{Lafreniere2010} measured its common proper motion over a $1~yr$ baseline and spectroscopically identified it as a young L4 spectral type. Unlike for GSC~06214B, the near-IR (NIR) spectrum of this object shows no sign of active accretion. At $8^{+4}_{-2}~M_{Jup}$ (based on $5~Myr$ DUSTY models), this object is below the deuterium-burning limit nominally separating brown dwarfs from planets \citep{Spiegel2011a}, although 
it has a higher mass ratio and much larger separation than directly-imaged planets such as $\beta$~Picb and HR~8799bcde \citep{Lagrange2010, Marois2008, Marois2010}.  Thus, for simplicity we refer to this object as a BD. 

\item HIP~78530A is a $\sim2.5~M_{\odot}$ B9 star \citep[and references therein]{Lafreniere2011}; its USco membership is based on direct parallax measurements by the Hipparcos survey \citep[$157\pm13~pc$,][]{VanLeeuwen2007}. HIP~78530B was discovered at a projected separation of $4.5''$, or $\sim700~AU$, by \citet{Lafreniere2011}. The authors confirmed its common proper motion over a $2~yr$ baseline and spectroscopically classified the BD as a low-surface gravity M8. They used $5~Myr$ DUSTY models to estimate a mass of $23\pm3~M_{Jup}$. As with 1RXS~1609B, this object's NIR spectrum shows no indication of active accretion.
\end{itemize}

\begin{figure}
\begin{centering}
%\epsscale{0.8}
%\plotone{fig1.eps}
\plotone{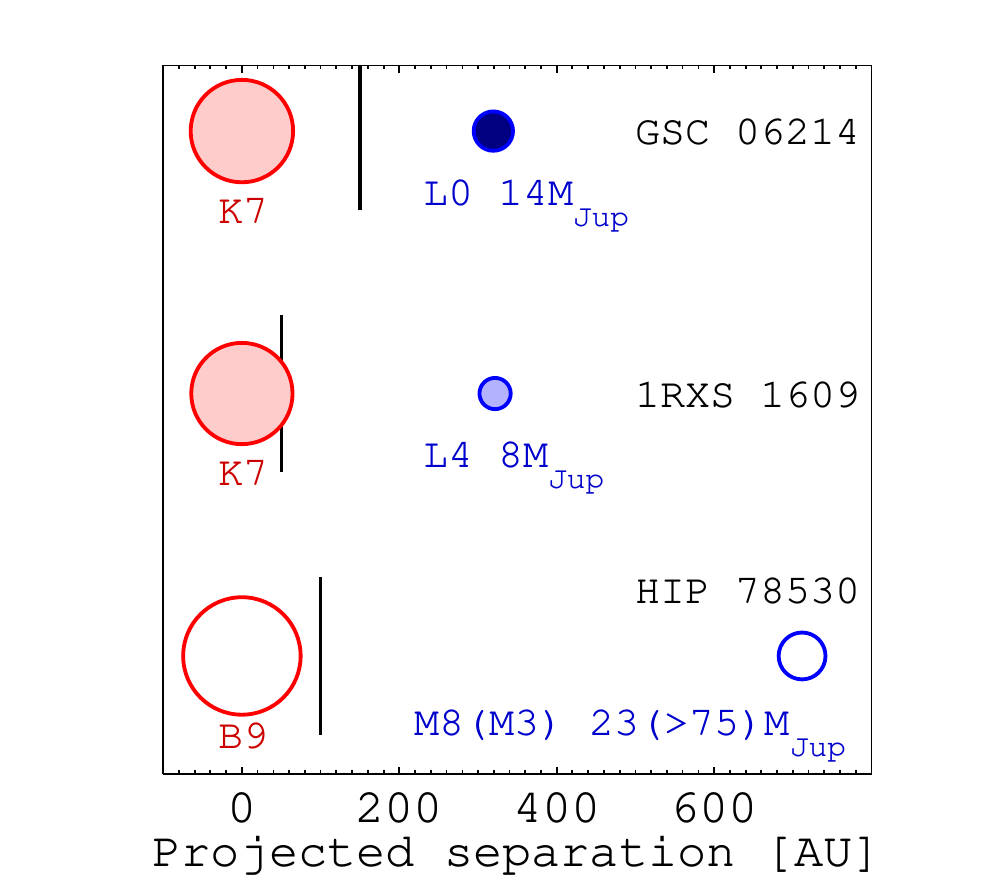}
\caption{Schematic diagram of the three USco system properties (discussed further in Sections \ref{sect:ResultsAnalysis} and \ref{sect:Discussion}). Circle diameters are proportional to the log of the object's mass. The black line is the projected separation beyond which objects as luminous as the ``B'' components would have been detected in the literature or our data. Fill indicates disk properties: objects without $1-5\um$ or $24\um$ excess are open circles; the dark fill of GSC~06214B signifies its $1-5\um$ excess and actively accreting disk; light shading refers to unresolved $24\um$ excess in the system which may be around the primary and/or companion.}
\label{fig:UScoSummary}
\end{centering}
\end{figure}

These three systems (Figure \ref{fig:UScoSummary}) may probe two very different epochs in stellar evolution for the primaries and different internal structures for the companions. The B star HIP~78530A is on or near the main sequence, while the K stars 1RXS~1609A and GSC~06214A 
 are still descending onto the main sequence. HIP~78530B is a BD whose mass is clearly above the deuterium-burning limit, while GSC~06214B is near, and 1RXS~1609B is likely below, this limit (for an assumed age of $5~Myr$).

\subsection{Observations}

For resolved $3-5\um$ observations of each system, we used the near/thermal-infrared imaging camera, Clio, at the $6.5~m$ MMT  and/or the Large Binocular Telescope Interferometer (LBTI) at the $8.4~m$ LBT (in single-sided, non-interferometric mode). A summary of these observations is presented in Table \ref{tab:obs}. At $24\um$, we used archival Spitzer/MIPS data for all three systems.   Because the MIPS point-spread function (PSF) full-width half-maximum (FWHM)  is $1.3-2.7$ times the projected separations of each companion from its host star, none of the systems are resolved at 24 $\mu m$.

\begin{deluxetable*}{llcccc}
\tablecaption{Summary of LMIRCam and Clio observations.\label{tab:obs}}
\tablewidth{0pt}
\tablehead{
\colhead{Object}  & \colhead{Date}  & \colhead{Instrument}  & \colhead{Filter}  & \colhead{Wavelength Range ($\mu m$)} & \colhead{Integration\ (min)}   
}  
\startdata
GSC~06214   & 2011 May 26  & LMIRCam   &  $M^\prime$   &	$4.60-4.97$  &   30  \\
            & 2011 Apr 22  & Clio      &  $3.1\um$    &   $3.03-3.16$	 &   45 \\
            & 2011 Apr 22  & Clio      &  $3.3\um$    &   $3.13-3.52$	&	20 \\
1RXS~1609   & 2011 Apr 17  & Clio      &  $3.1\um$    &   $3.03-3.16$	 &   45 \\
            & 2011 Apr 17  & Clio      &  $3.3\um$    &   $3.13-3.52$	&	30 \\
HIP~78530   & 2011 May 26  & LMIRCam   &  $L^\prime$   &  $3.42-4.12$	& 15  \\
\enddata
\end{deluxetable*}

\subsubsection{LBT/LBTI/LMIRCam $L^\prime$ and $M^\prime$ Imaging of GSC~06214 and HIP~78530}
\label{sect:LBT/LBTIImaging}

The L and M Infrared Camera (LMIRCam) is the $3-5\um$ channel of LBTI \citep{Hinz2008, Skrutskie2010}. The field of view with an $f/15$ beam is $11\arcsec$ square, with $10.7~mas$ pixels (calibrated Nov.\ 2011 by \citet{Rodigas2012}). We assume the instrument rotation angle is repeatable to $1\dgr$.  Field distortion has not been measured observationally; ZEMAX models of the LMIRCam optics predict field distortions of $<2\%$ over distances $<5''$.

The LBT AO system employs a deformable secondary mirror in order to minimize the number of warm surfaces along the optical path. As a result, it benefits from increased sensitivity in the thermal IR \citep{Lloyd-Hart2000}. For system details see \citet{Esposito2011} (E11). The LBTI AO hardware design is identical to that used in E11, but the calibration is independent. We corrected 200 aberration modes, rather than the 400 modes E11 used to achieve $>80\%$ Strehl ratio (SR) in H band. At the time of these observations, the LMIRCam PSF was degraded due to residual instrumental astigmatism (which has since been mitigated). Nonetheless, because wavefront phase errors decrease with increasing wavelength, our observations approached the diffraction limit. During this run, we achieved a SR on LMIRCam of $\sim80\%$ at $M^\prime$ on $\alpha$~CrB ($R=2.2$) at airmass $\sim1$.  Data recorded directly from the wavefront sensor itself, not subject to the LMIRCam instrumental astigmatism, indicate the AO system was delivering $\sim90\%$ SR during these observations. The science observations were executed under more difficult conditions: low elevation (high airmass), increased winds, and with fainter guide stars, and so were subject to poorer AO performance than the $\alpha$~CrB observations. Thus the sensitivities we derive in this paper are somewhat lower than those which can now be achieved with LBTI on more favorable targets \citep[e.g.][]{Rodigas2012, Skemer2012} or with its recently improved AO calibrations.

We observed GSC~06214 and HIP~78530 on UT 2011 May 26. During this LBTI commissioning run, only the AO system on the right LBT aperture was fully operational, so we observed in single aperture mode. The observing strategy for these, and all other LBTI observations, was Angular Differential Imaging (ADI) \citep{Marois2006}: the instrument rotation angle was fixed, and the field was allowed to rotate on the detector over the course of the observations. We used a four position nod pattern for taking image data, with nods of $1.5-3''.$ Due to the high sky background, we were limited to very short exposures: $87~ms$ at $M^\prime$ and $146~ms$ at $L^\prime$. To limit data volume, 200 individual exposures were coadded per single saved image. We obtained $30~min$ of $M^\prime$ data on GSC~06214 and $15~min$ of $L^\prime$ data on HIP~78530. The FWHM (average of major and minor axes of the astigmatic PSF) for the HIP~78530 ($L^\prime$) and GSC~06214 ($M^\prime$) images were $0.12''$ and $0.18''$, respectively.

\subsubsection{MMT/Clio 3.1$\um$ and 3.3$\um$ Narrowband Imaging of GSC~06214 and 1RXS~1609}
\label{sect:MMT/ClioImaging}

Clio is also a narrow-field $3-5\um$ imager \citep{Sivanandam2006}. At the time of these observations it was installed on MMT; it is now at the Magellan Clay telescope. The MMT AO system has an architecture similar to that of LBT AO, with a deformable secondary mirror correcting 52 aberration modes \citep{Brusa2004}. In $f/15$ mode, the Clio plate scale is $29.9\pm0.1~mas~pixel^{-1}$, and the field of view of the entire array is $15\arcsec\times30\arcsec$ (calibrated July 2010 using the binary system HIP~88817). We assume the instrument rotation angle is repeatable to $0.1\dgr$.  Plate scale distortion has not been measured, but ZEMAX models predict distortion of $<0.2\%$ over 5'' separations. 

We observed 1RXS~1609 and GSC~06214 on UT 2011 April 17 and 22, respectively. In order to decrease time lost to detector readout, we used only a subarray with a field of view of $9\arcsec\times30\arcsec.$ Integrations were 10 coadds of $5-6~s$ at $3.1\um$ and 40 coadds of $1~s$ at $3.3\um$. The data were taken in ADI mode in a $15''$ two-position nod, with $1-2''$ dithers between nods. These observations suffered from moderate wind from the South that induced vibrations in the telescope and caused smearing in some of the images. The resultant FWHM ranged from $0.20''-0.35''$, depending on the severity of the vibrations. On GSC~06214, we obtained approximately $45~min$ and $20~min$ of on-source integration in the $3.1\um$ and $3.3\um$ narrowband filters, respectively. On 1RXS~1609 we obtained $45~min$ of integration at $3.1\um$ and $30~min$ at $3.3\um$.

\subsubsection{Spitzer/MIPS $24\um$ Imaging}

Each system had been previously observed by MIPS.  1RXS~1609 and HIP~78530 $24\um$ photometry are published in \citet{Carpenter2009}, and we refer the reader to this publication for a description of the observations.  For GSC~06214, we analyzed archival MIPS data taken as part of the Spitzer Legacy program: Gould's Belt (PID 30574, PI: L. Allen), which aimed to provide a complete census of star-forming clouds in the solar neighborhood. GSC~06214 was imaged in two AORs: 20000768 and 20001280, obtained with the fast scan rate ($3~s$ per frame). The system was located near the end of the scan map; therefore, only 5 frames per pointing were obtained near the source position with an equivalent integration time of $15~s$ at $24\um$. We detected the source only at $24\um$ due to the shallowness of the survey;  we found no source in the corresponding 70 and $160\um$ maps. Therefore, we only discuss the $24\um$ result hereafter. Multiple frames were mosaicked based on the image WCS pointing information and subsampled by a factor of 2, resulting in a plate scale of 1.245\arcsec$~pixel^{-1}$.

\subsection{Image Processing}
\label{sect:ImageProcessing}

\subsubsection{LMIRCam and Clio Data}
\label{sect:LCImageProcessing}
Custom MATLAB scripts were used for reducing the LBTI and Clio data. Background, bias, and dark subtraction were achieved by differencing images from adjacent nods. Bad pixels were flagged using a nearest-neighbors median filter and replaced with that median. In LMIRCam, each of the channels had a small time-varying bias level offset as well as variable pattern noise. We subtracted the median value of each channel to remove the offset. Then we subtracted a 2-D template of the channel read noise created from the median of all channels. For those calculations we masked out a $\sim2''$ square region containing the source to avoid biasing the median. Finally, we aligned the images according to their position angle, rotating so North was up. 

These observations did not achieve sufficient on-sky rotation to synthesize and subtract the primary's PSF, as per \citet{Marois2006}. However all secondaries were sufficiently separated from their primaries ($>2\arcsec$) that contamination from the primary was negligible. To produce the final image, we mean-combined our set of rotated images together, and $3-4\sigma$ outliers for each pixel in the image stack were clipped.

Aperture photometry was used to extract the differential flux between primary and secondary. None of the objects were resolved, so we used aperture radii of $\sim1$~FWHM. The background was estimated using neighboring apertures of the same radius at the same separation from the host star. No aperture correction was necessary, because the same aperture radius was used for both the primary and secondary. 

We next calculated sensitivity as a function of radial separation from the primary star. Residual low-spatial frequency offsets were removed by subtracting a smoothed image (Gaussian FWHM=$10\lambda/D$) from the science image. The resulting image was then smoothed with a disk the same diameter as the aperture used for photometry. The noise, $\sigma$, as a function of separation was calculated by measuring the standard deviation of 1 pixel wide annuli centered on the primary. From this we calculated the signal to noise ratios (SNR) of our detections. The uncertainty in our photometry (in magnitudes) is given by $2.5\cdot(ln(10)\cdot SNR)^{-1}$. The noise counts as a function of radius were also converted to contrast (delta magnitudes between the primary star and the measured quantity, noise in this case) by comparing to the peak pixel value of the primary star in the smoothed image. 

The primary star photometry in our filters was estimated by interpolating between published photometry in adjacent filters (i.e.: between shorter and longer wavelength photometry), because no standard stars were observed.  There are no strong spectral features in K7 or B9 stars between $2\um$ and $5\um$ which would bias this interpolation. The 2MASS and WISE surveys recorded unresolved photometry for each system. Additionally, \citet{Carpenter2006a} conducted a Spitzer IRAC survey which included 1RXS~1609 and HIP~78530. As we will show, the BD contribution to the system flux at $1-5\um$ is negligible, so we can treat 2MASS, WISE W1 and W2, and Spitzer IRAC photometry as measurements of the primary star only. Where available, we also used resolved photometry at additional wavelengths from previous work. The data used for interpolation are listed in Tables \ref{tab:GSC}, \ref{tab:1RXS}, and \ref{tab:HIP}.

\subsubsection{Spitzer MIPS Data}
We reduced the raw $24\um$ data of GSC~06214 using the MIPS Data Analysis Tool \citep{Gordon2005} with post-pipeline processing to improve flat fielding and remove instrumental artifacts as detailed in \citet{Engelbracht2007}. We extracted the photometry using PSF fitting with the IDL-based {\it StarFinder} program \citep{Diolaiti2000}. The input PSF was constructed using observed stars and a smoothed STINYTIM model PSF; it has been tested to ensure the photometry results are consistent with the MIPS calibration \citep{Engelbracht2007}.

%***************************************************************************************

\section{Results and Analysis}
\label{sect:ResultsAnalysis}

In this section, we present photometry and analysis for each of the three binary systems. We first measure photometry for both components, from both our newly obtained high-resolution near-IR imaging and archival unresolved Spitzer $24\um$ imaging. Interstellar extinction/reddening can mimic the red excess from a hot disk component; the average extinction towards USco is small ($A_V<1$), but not negligible, and we estimate it as necessary. We re-examine the companions' spectral types and constrain their near-IR (hot dust) excesses based on their placement in color-color diagrams compared to similarly young low-mass objects. We use unresolved Spitzer $24\um$ data to probe the warm dust in the system as a whole (both A and B components together). Finally, we quantify our sensitivity to additional objects in the system at intermediate separations from the primary.

\subsection{GSC~06214}

\subsubsection{Photometry and Astrometry}
\label{sect:GSCPhotometry}

\begin{figure*}
\begin{centering}
%\plotone{fig2.eps}
\plotone{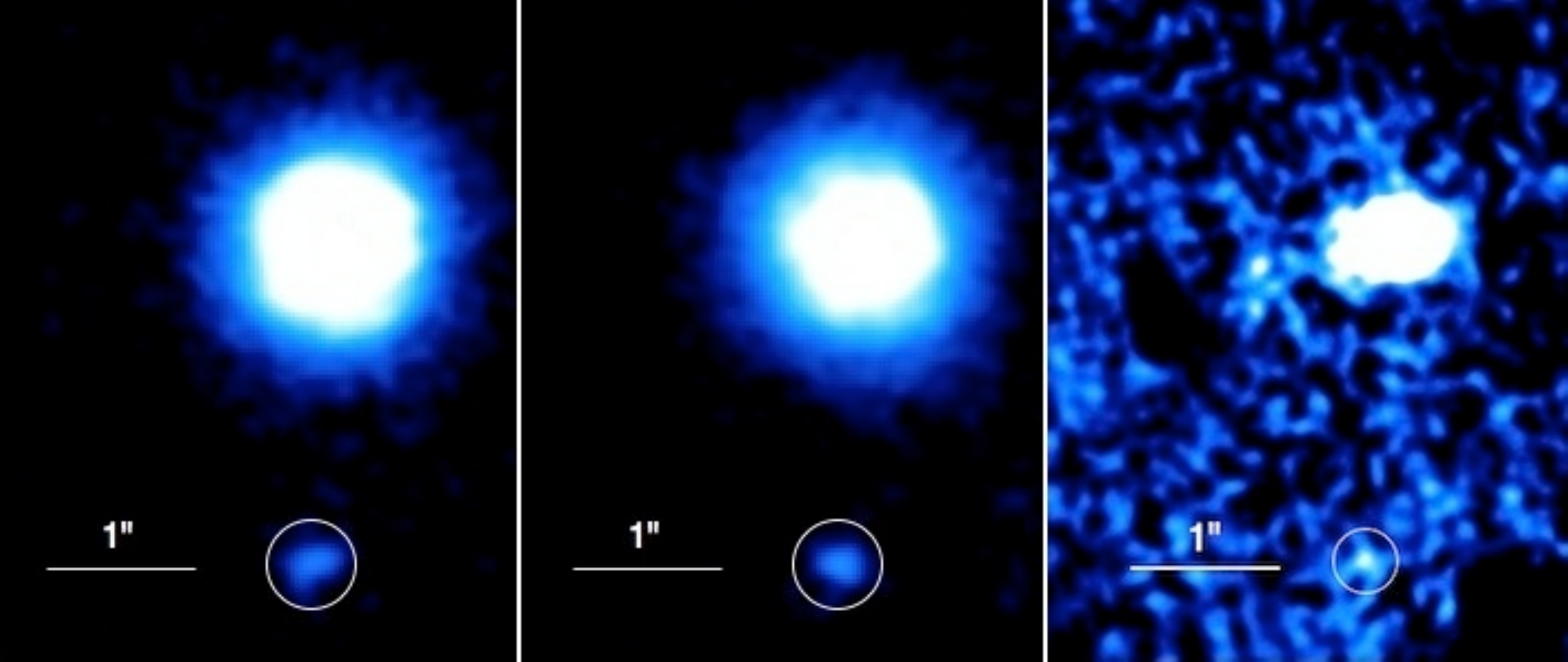}
\caption{GSC~06214 Clio and LBTI images. \textit{Left:} Clio $3.1\um$. \textit{Center:} Clio $3.3\um$. \textit{Right:} LBTI $M^\prime$. The companion, circled, is $2.2\arcsec$ to the south of the primary. For display purposes, all images have been smoothed with a Gaussian kernel with FWHM=$1\lambda/D\sim0.1''$. In the $M^\prime$ image, the bright spot to the left of the primary is an artifact; the dark spots to the left and bottom right are residuals from nod subtraction.}
\label{fig:GSC}
\end{centering}
\end{figure*}

The final Clio images at $3.1\um$ and $3.3\um$ are shown in Figure \ref{fig:GSC}. At $3.1\um$ we measured a contrast between ``A'' and ``B'' of 5.26 magnitudes, and at $3.3\um$ we measured a contrast of 5.30 magnitudes. Using the procedure described in Section \ref{sect:LCImageProcessing}, we inferred the $3.1\um$ and $3.3\um$ magnitudes of the primary to be $9.11\pm0.05$ and $9.10\pm0.05$, respectively.  The $5\sigma$ sensitivity limits for these two filters were magnitudes 15.82 and 16.14, yielding signal to noise ratios of 18 and 23 for the detections in each wavelength. From this we derive, $[3.1]_B = 14.41\pm0.08$ and $[3.3]_B = 14.40\pm0.07$ (Table \ref{tab:GSC}). 

We calculated a separation of 2.17$\pm$0.02\arcsec and a position angle of 175.9$\pm$0.1$\dgr$, consistent at the $1.5\sigma$ level with astrometry measured by \citet{Ireland2011}. Because the companion was not detected in individual frames, we estimated the uncertainty by comparing the values calculated from the final images at each wavelength. Our quoted uncertainty accounts for this comparison as well as our assumed uncertainties in calibration and distortion (see Section \ref{sect:MMT/ClioImaging}).

The final LMIRCam $M^\prime$ image is also presented in Figure \ref{fig:GSC}. The BD was detected at a contrast of 4.65 magnitudes. For the primary, we estimated $M^\prime_A$ = 9.10$\pm$0.05.  Based on the $5\sigma$ sensitivity limit of $13.39~mag$, the BD detection has $SNR=3.6$, and therefore $M^\prime_B$ = 13.75$\pm$0.3.  Even though this SNR is lower than that nominally adopted to identify candidate companions in high-contrast imaging data \citep[$5-6\sigma$;][]{Marois2008}, we also detect point sources at the same location in the $3.1\um$ and $3.3\um$ images, justifying a lower detection threshold. The bright spot $\sim1\arcsec$ to the east of the primary does not appear in the more sensitive $3.1$ and $3.3\um$ images and is almost certainly an artifact.

The $24\um$ flux density of the system is $3.3\pm0.3~mJy$ ($8.34\pm0.1~mag$); the error is dominated by shot noise.

\begin{deluxetable}{lll}
\tablecaption{GSC~06214-00210 system properties and apparent photometry.\label{tab:GSC}}
\tablewidth{0pt}
\tablehead{
\colhead{Property}	& \colhead{GSC~06214A} 	& \colhead{GSC~06214B}
}  
\startdata
Distance $[pc]$ \tablenotemark{a}    &\multicolumn{2}{c}{145$\pm$20}    \\
Spectral type \tablenotemark{b}   & K7$\pm$0.5    & L0$\pm$1              \\
Mass \tablenotemark{b}	           & 0.9$\pm$0.1 \msun & 14$\pm$2 $M_{Jup}$  \\
$T_{eff}~[K]$ \tablenotemark{b}	   & 4200$\pm$150   &	2200$\pm$100	      \\
$log(L/L_\odot)$ \tablenotemark{b}  &  -0.42$\pm$0.08        &   -3.1$\pm$0.1     \\
Separation $['']$     \tablenotemark{c} &   \multicolumn{2}{c}{2.17$\pm$0.02}    \\
PA $[\dgr]$  	      \tablenotemark{c} &   \multicolumn{2}{c}{175.9$\pm$0.1}	    \\
$J$    & 9.998$\pm$0.027 \tablenotemark{d}  & 16.24$\pm$0.04 \tablenotemark{e}    \\
$H$    & 9.342$\pm$0.024 \tablenotemark{d}  & 15.55$\pm$0.04 \tablenotemark{e} 	\\
$K$    & 9.152$\pm$0.021 \tablenotemark{d}  & 14.95$\pm$0.05 \tablenotemark{e} 	\\
$3.1\um$  \tablenotemark{c} & 9.11$\pm$0.05  &   14.41$\pm$0.08 \\
$3.3\um$  \tablenotemark{c}  & 9.10$\pm$0.05  &   14.40$\pm$0.07 \\
$3.4\um$  \tablenotemark{f}  & \multicolumn{2}{c}{9.083$\pm$0.022}    \\
$L^\prime$ \tablenotemark{e}   & 9.10$\pm$0.05  &   13.75$\pm$0.07 \\
$4.6\um$  \tablenotemark{f}  & \multicolumn{2}{c}{9.107$\pm$0.020}  \\
$M^\prime$  \tablenotemark{c} & 9.10$\pm$0.05  &   13.75$\pm$0.3 \\
$12\um$  \tablenotemark{f}   & \multicolumn{2}{c}{8.964$\pm$0.032}   \\
$24\um$    \tablenotemark{c} & \multicolumn{2}{c}{8.34$\pm$0.1} \\
\enddata

\tablenotetext{a}{Mean cluster distance \citep{DeZeeuw1999, Preibisch2002}.}
\tablenotetext{b}{\citet{Bowler2011}.}
\tablenotetext{c}{This work.}
\tablenotetext{d}{2MASS $J/H/K_S$ survey \citep{Skrutskie2006}.}
\tablenotetext{e}{\citet{Ireland2011}. MKO $J/H/K^\prime/L^\prime$.}
\tablenotetext{f}{System. WISE survey \citep{Wright2010}.}

\end{deluxetable}

\subsubsection{Extinction Estimate}
\label{sect:extinction}
 
\begin{figure}
\begin{centering}
\plotone{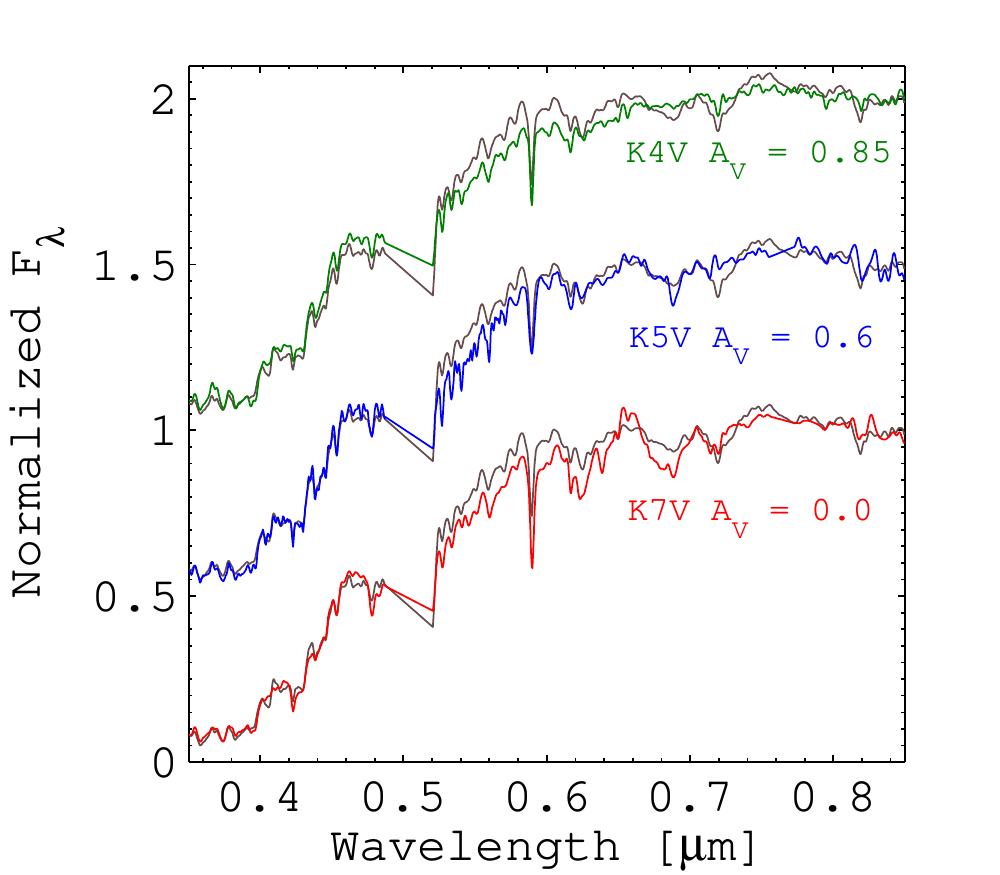}
%\plotone{fig3.eps}
\caption{GSC~06214A flux vs.\ wavelength for best fit spectral type and extinction. $F_\lambda$ is offset vertically for clarity. Gray is the stellar spectrum \protect\citep{Bowler2011}, red is a K7V with no extinction, blue is a K5V with $A_V=0.6~mag$, and green is a K4V with $A_V=0.85~mag$. We find $A_V\leq0.6~mag$ for the primary.}
\label{fig:GSCA_Spec}
\end{centering}
\end{figure}

GSC~06214A did not have a published extinction value, so we performed our own estimate. We found $A_V\leq0.6$ using the following prescription. We compared the primary's flux-calibrated optical spectrum \citep[courtesy][]{Bowler2011} with \citet{Pickles1998} spectral standards (Figure \ref{fig:GSCA_Spec}). The observed and template spectra were normalized by their $800-900~nm$ averages. Interstellar extinction curves with $R_V = 3.1$ \citep{Cardelli1989} were used. Based on $\chi^2$ fitting, both K7V with $A_V=0$ and K5V with $A_V=0.6$ provide acceptable fits to the data. K4V and $A_V=0.85$ is a worse fit to both the continuum and spectral features. In order to make the most conservative estimate of the IR excess in the system, we adopt the maximum allowed extinction value: $A_V=0.6~mag$. 

\subsubsection{Spectral Classification and Evidence for a Near-IR Excess}
\label{sect:color-color}

We placed GSC~06214 in infrared color-color diagrams, compared to several observational brown dwarf and planet samples: field brown dwarfs \citep{Leggett2010}, very young substellar objects (L2010), the young planetary-mass object 2M~1207b \citep{Chauvin2004, Mohanty2007, Skemer2011, Barman2011}, and the planets HR~8799bcd \citep{Marois2008a, Marois2010, Currie2011b, Galicher2011, Skemer2012}. The L2010 sample of M0$-$L0 dwarfs was derived from dereddened observations of Taurus, Chameleon, and TW Hya, so it provides the most appropriate comparison sample for our similarly young brown dwarfs.  

\begin{figure*}
\begin{centering}
\plottwo{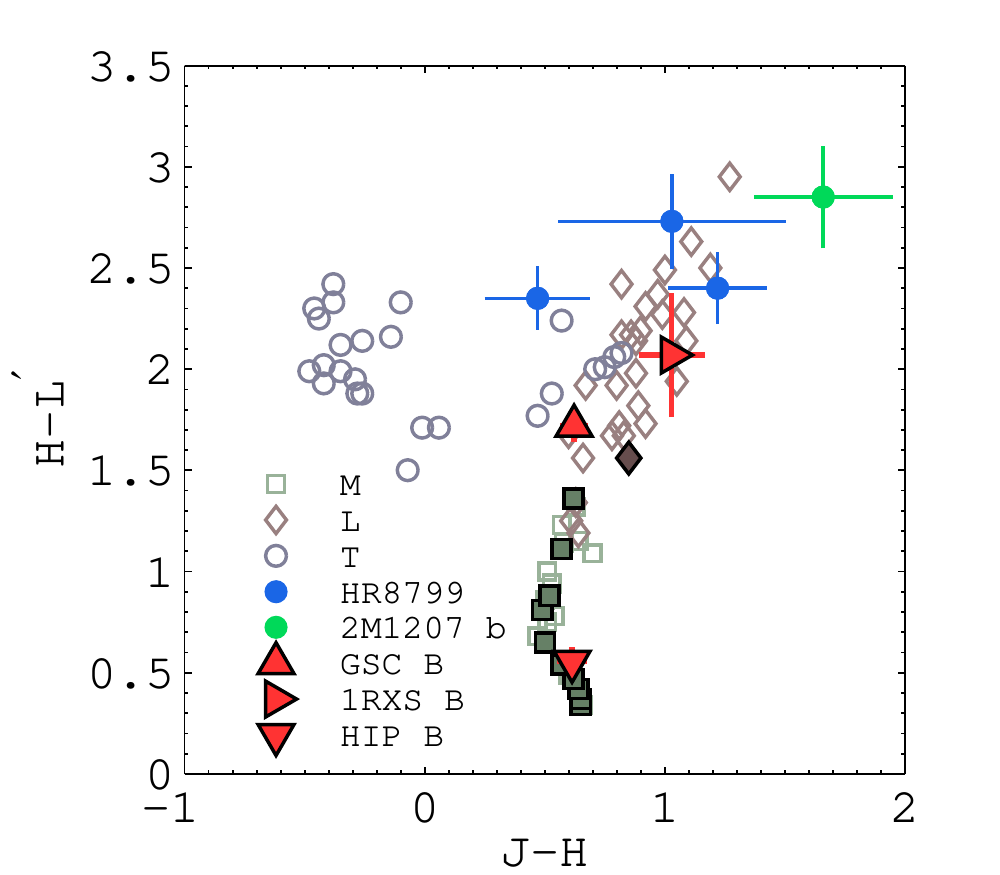}{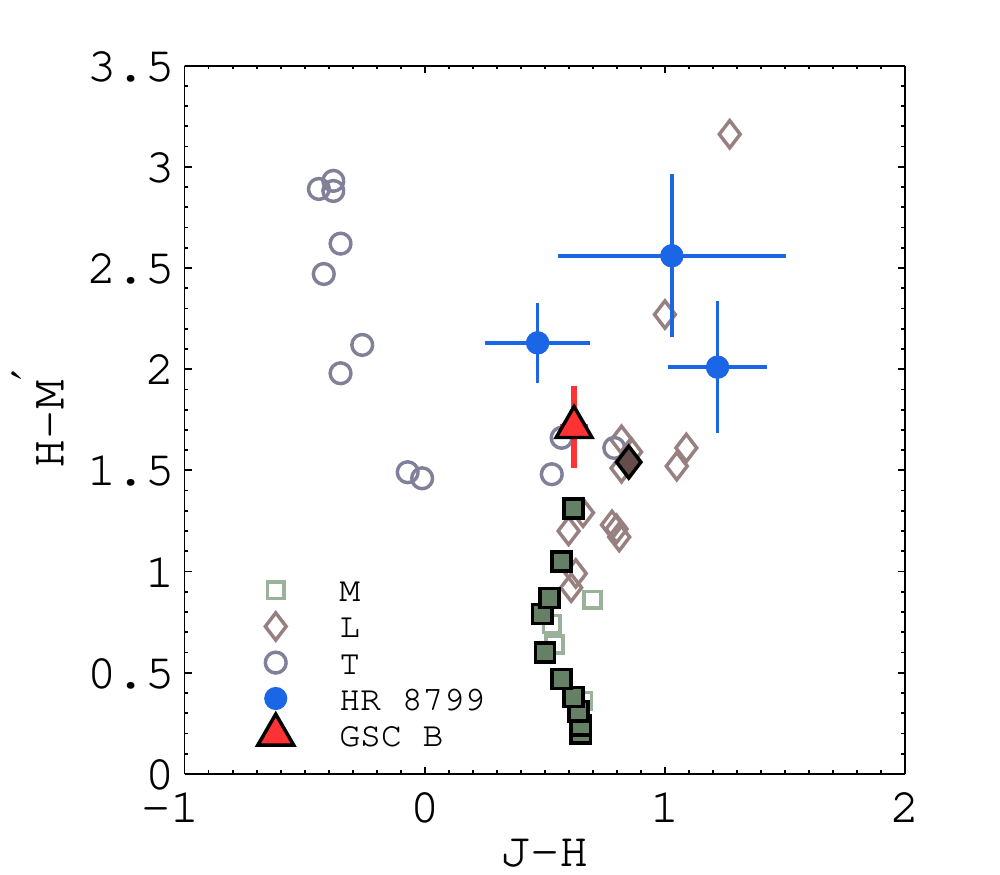}
%\plottwo{fig4a.eps}{fig4b.eps}
\caption{$H-L^\prime$ vs.\ $J-H$ \textit{(left)} and $H-M^\prime$ vs.\ $J-H$ \textit{(right)} color-color diagrams  of dereddened USco companions and comparison samples. Open green squares, brown diamonds, and gray circles: field M, L, and T dwarfs from \protect\citet{Leggett2010}; filled green squares and brown diamond: young M0--L0 dwarfs from L2010; filled blue circles: young planets HR~8799bcd \protect\citep{Currie2011b, Marois2008, Marois2010, Skemer2012, Galicher2011}; filled green circle: young planet 2M~1207b \protect\citep{Mohanty2007}.}
\label{fig:colorcolor1}
\end{centering}
\end{figure*}

In order to conduct a homogeneous analysis, we transformed all previously published $J$, $H$, and $K$ BD photometry to the MKO $J/H/K$ system. For 2MASS $J/H/K_S \rightarrow$ MKO $J/H/K$, we used the transformations published in \citet{Leggett2006}. For MKO $K^\prime \rightarrow K$ of stars, we used \citet{Wainscoat1992}; however, we could not apply this transformation to the BD, because it is only calibrated for $H-K < 0.4$. Therefore, we calculated the $K^\prime \rightarrow K$ transformations by integrating over DUSTY evolutionary model spectral energy distributions (SEDs) \citep{Chabrier2000}.  We took the SED that most closely matched the color of the BD in our filters of interest (in this case, $H-K$). We used the same procedure for all color transformations for which empirical relations were not available. We calculated $K=14.75~mag$ for GSC~06214B. We then dereddened all GSC~06214B photometry, using the extinction value for the primary ($A_V=0.6$) discussed in Section \ref{sect:extinction}. We presumed the extinction toward the primary and secondary was the same, a nominally valid assumption, though counterexamples do exist \citep[e.g. GG Tau,][]{White2001}. Figure \ref{fig:colorcolor1} shows the color-color diagrams $H-L^\prime/J-H$ and $H-M^\prime/J-H$.

GSC~06214B was previously classified using $J/H/K$ spectra as a spectral type of L0$\pm$1 \citep{Bowler2011}. Its $J-H$ color is consistent with an L2010 spectral type of M9. The excesses at $K$, $L^\prime$, and $M^\prime$ compared to an L2010 M9 dwarf photosphere are $0.15\pm0.05$, $0.35\pm0.07$, and $0.4\pm0.3~mag$, respectively, accounting for photometric errors only. Fortuitously, the measured excesses are similar even if we use $A_V=0$, because reddening moves the object up and right in color-color space parallel to the line connecting L2010 M9 and L0 spectral types. Therefore, our conclusions about the color excess for this object are independent of the extinction estimate, over the allowed range of extinction values.

\subsubsection{$24\um$ Flux}
\label{sect:GSC24umexcess}

The GSC~06214 system was imaged with Spitzer/MIPS at $24\um$. The PSF is broader than the host-companion separation, so it is not possible to measure the primary and secondary objects' emission independently. However, the BD's photosphere does not contribute significantly at this wavelength. Whether based on a Rayleigh-Jeans extrapolation from short wavelength data or using DUSTY model SEDs, the BD's photospheric emission is expected to be less than 1\% of the observed $24\um$ flux.

We compared the MIPS photometry to the $24\um$ flux predicted for the ``A'' component by an empirical polynomial relation between $V-K_S$ and $24\um$ for main sequence stars \citep{Urban2012}. The $24\um$ flux prediction is robust to order $0.01~mag$ between early B and early M spectral types. GSC~06214A is likely in the final stages of contraction onto the main sequence. However, the variation in $V-K_S$ color, and therefore in inferred $24\um$ flux, between this object and a main sequence star, is less than the combined observational uncertainties at $V,~K_S,$ and $24\um$. 

From this comparison, we find evidence of a $24\um$ excess in the system. The observed system flux is $3.3\pm0.3~mJy$, while the predicted flux from ``A'' is $1.73\pm0.15~mJy$. This translates to an excess of $91\pm24\%$ above the expected photospheric emission from ``A.'' The uncertainty is dominated by the low signal to noise of the $24\um$ data. Our detection corroborates the $2\sigma$ excess at WISE $22\um$ found by \citet{Bowler2011}. Therefore, one or both members of the system must retain a warm excess.

\subsubsection{Constraints on additional objects}
\label{sect:GSCcompanions}

No additional objects are detected in our observations. As discussed in Section \ref{sect:GSCPhotometry}, the candidate at $~1''$ separation in $M^\prime$ is likely an artifact. In our collection of images, we could have detected additional companions of mass $\gtrsim20~M_{Jup}$ at separations $\gtrsim150~AU$ from the primary, with a limit of $\sim10-12~M_{Jup}$ beyond $200~AU$. Previous aperture masking and direct imaging work have placed deeper limits of $35/15/5~M_{Jup}$ at $10/150/300~AU$ from the primary \citep{Kraus2008,Ireland2011}. We would have resolved an equally bright binary companion to ``B'' (i.e. ``Bb'')  if it was $>20~AU$ projected separation from ``B.''

%%%%%%%%%%%%%%%%%%%%%%%%%%%%%%%%%%

\subsection{1RXS 1609}

\subsubsection{Photometry and Astrometry}
\label{sect:1RXS1609}

\begin{figure*}
\begin{centering}
\epsscale{0.9}
%\plottwo{fig5a.eps}{fig5b.eps}
\plottwo{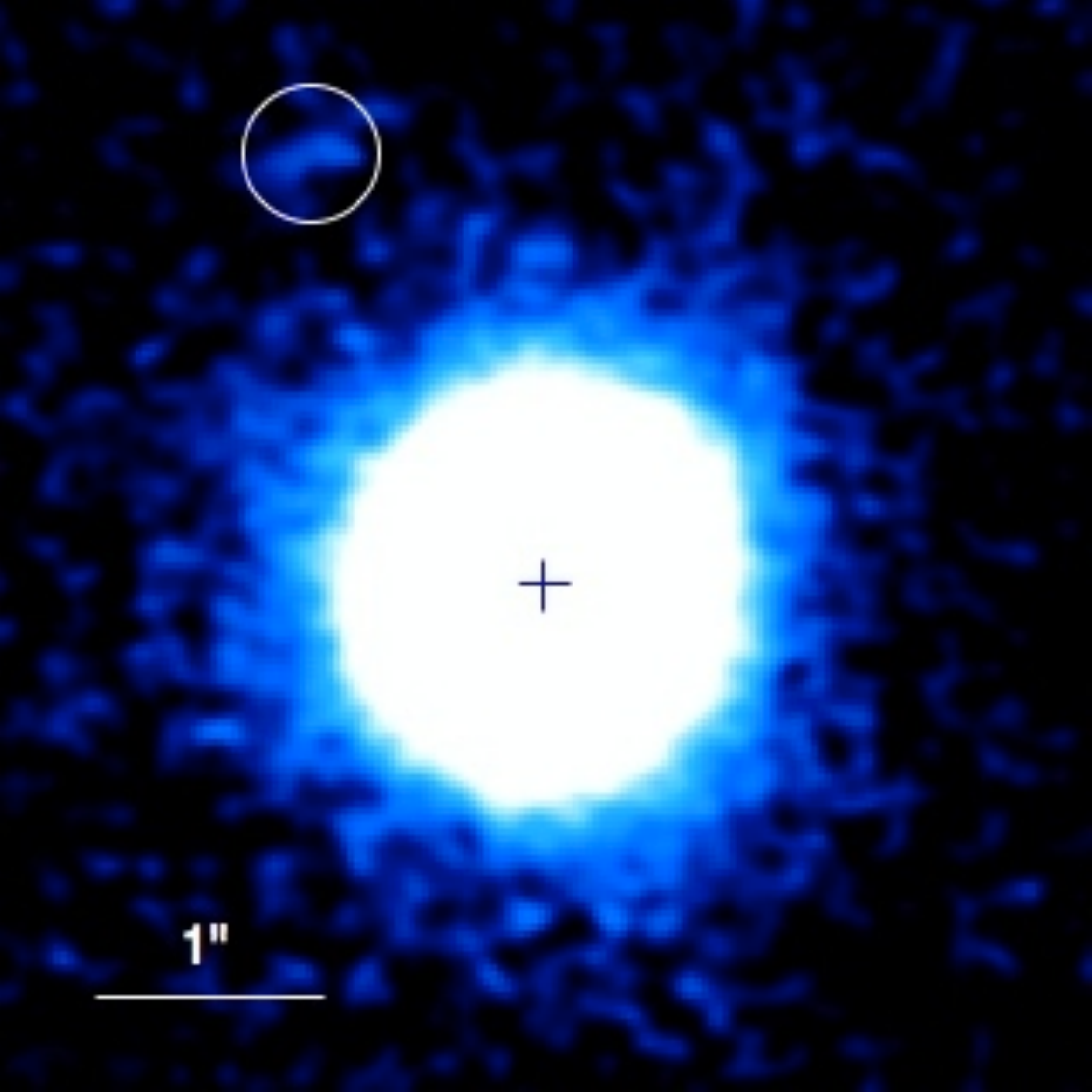}{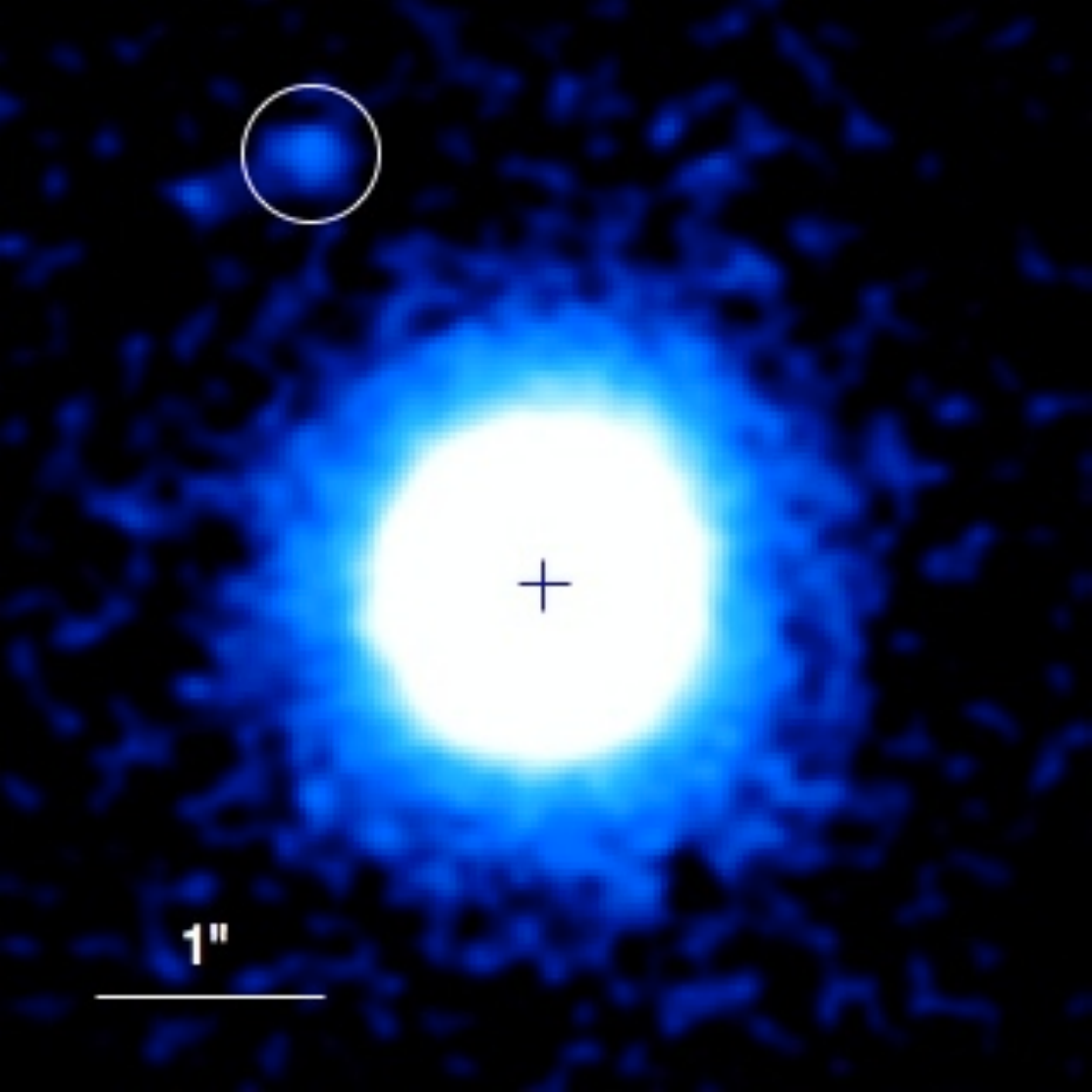}
\caption{1RXS~1609 Clio images. \textit{Left:} $3.1\um$. \textit{Right:} $3.3\um$. For display purposes, both are smoothed with a Gaussian kernel with FWHM=$1\lambda/D\sim0.1''$. The companion, circled, is $2.2\arcsec$ to the NE.}
\label{fig:1RXS3133}
\end{centering}
\end{figure*}

Figure \ref{fig:1RXS3133} shows the final 1RXS~1609 narrowband images. We estimated the primary's $3.1\um$ and $3.3\um$ magnitudes to be $8.80\pm0.05$ and $8.78\pm0.05$, respectively. At $3.1\um$ we found a contrast between ``A'' and ``B'' of $6.85~mag$. The $5\sigma$ background sensitivity limit of $15.71~mag$ yielded $SNR=5.3$ for the detection, thus $[3.1]_B= 15.65\pm0.21$. At $3.3\um$ we found $6.4~mag$ contrast, a sensitivity limit of $16.14~mag$, $SNR=7.5$, and $[3.3]_B=15.2\pm0.16$. (See Table \ref{tab:1RXS}).  

We measured a separation of $2.15\arcsec\pm0.03\arcsec$ at $PA=28.3\dgr\pm0.4\dgr$, consistent with \citet{Lafreniere2010} at the $2\sigma$ level. These values represent the average of the astrometry calculated at $3.1\um$ and $3.3\um$. Errors account for measurement uncertainties as well as our assumed calibration and distortion uncertainties (see Section \ref{sect:MMT/ClioImaging}).

\begin{deluxetable}{lll}
\tablecaption{1RXS~160929.1-210524 system properties and apparent photometry. \label{tab:1RXS}}
\tablewidth{0pt}
\tablehead{
\colhead{Property}	& \colhead{1RXS~1609A} 	& \colhead{1RXS~1609B}  
}
\startdata
Distance $[pc]$ \tablenotemark{a}  &\multicolumn{2}{c}{145$\pm$20}    \\
Spectral type  \tablenotemark{b}   & K7V   	  	& L$4^{+1}_{-2}$           \\
Mass           \tablenotemark{b}   & $0.85^{+0.20}_{-0.10}$\msun	& $8^{+4}_{-2}M_{Jup}$ \\
$T_{eff}$       \tablenotemark{b}   & $4060^{+300}_{-200}$	& $1800^{+200}_{-100}$ \\
$log(L/L_\odot)$ \tablenotemark{b}  & -0.37$\pm$0.15 & -3.55$\pm$0.2     \\
Separation $['']$ \tablenotemark{c} &   \multicolumn{2}{c}{2.15$\pm$0.03}    \\
PA $[\dgr]$	\tablenotemark{c}  &   \multicolumn{2}{c}{28.3$\pm$0.4}	\\
$J$ 	& 9.820$\pm$0.027 \tablenotemark{d}	& 17.90$\pm$0.12 \tablenotemark{b,e}   \\
$H$ 	& 9.121$\pm$0.023 \tablenotemark{d}	& 16.87$\pm$0.07 \tablenotemark{b,e}   \\
$K$ 	& 8.916$\pm$0.021 \tablenotemark{d}	& 16.19$\pm$0.05 \tablenotemark{b,e}   \\
$3.1\um$ \tablenotemark{c} & 8.80$\pm$0.05 	&  15.65$\pm$0.21      \\
$3.3\um$ \tablenotemark{c} & 8.78$\pm$0.05 	&  15.2$\pm$0.16       \\
$3.4\um$  \tablenotemark{g}  & \multicolumn{2}{c}{8.767$\pm$0.023}    \\
$L^\prime$ \tablenotemark{e,f}  & 8.73$\pm$0.05      & 14.8$\pm$0.3     \\
$4.5\um$ \tablenotemark{h}   & \multicolumn{2}{c}{8.80$\pm$0.01}      \\
$4.6\um$ \tablenotemark{g}   & \multicolumn{2}{c}{8.779$\pm$0.021}    \\
$8.0\um$ \tablenotemark{h}   & \multicolumn{2}{c}{8.735$\pm$0.005}    \\
$12\um$  \tablenotemark{g}   & \multicolumn{2}{c}{8.715$\pm$0.027}    \\
$24\um$ \tablenotemark{i}    & \multicolumn{2}{c}{8.43$\pm$0.01}      \\
\enddata

\tablenotetext{a}{Mean cluster distance \citep{DeZeeuw1999, Preibisch2002}.}
\tablenotetext{b}{\citet{Lafreniere2008}.}
\tablenotetext{c}{This work.}
\tablenotetext{d}{2MASS $J/H/K_S$ survey \citep{Skrutskie2006}.}
\tablenotetext{e}{MKO $J/H/K^\prime/L^\prime$.} 
\tablenotetext{f}{\citet{Lafreniere2010}}
\tablenotetext{g}{System. WISE survey \citep{Wright2010}.}
\tablenotetext{h}{System. Spitzer IRAC \citep{Carpenter2006a}.}
\tablenotetext{i}{System. Spitzer MIPS \citep{Carpenter2009}.}

\end{deluxetable}

\subsubsection{Spectral Classification and Near-IR Colors}

There is no extinction towards the 1RXS~1609 system \citep{Carpenter2009}. We transformed the $K^\prime$ companion photometry to yield $K=15.99\pm0.1$, following the procedure described in Section \ref{sect:color-color}. 1RXS~1609B is shown in the color-color diagrams in Figure \ref{fig:colorcolor1}.

\begin{figure}
\begin{centering}
%\plotone{fig6.eps}
\plotone{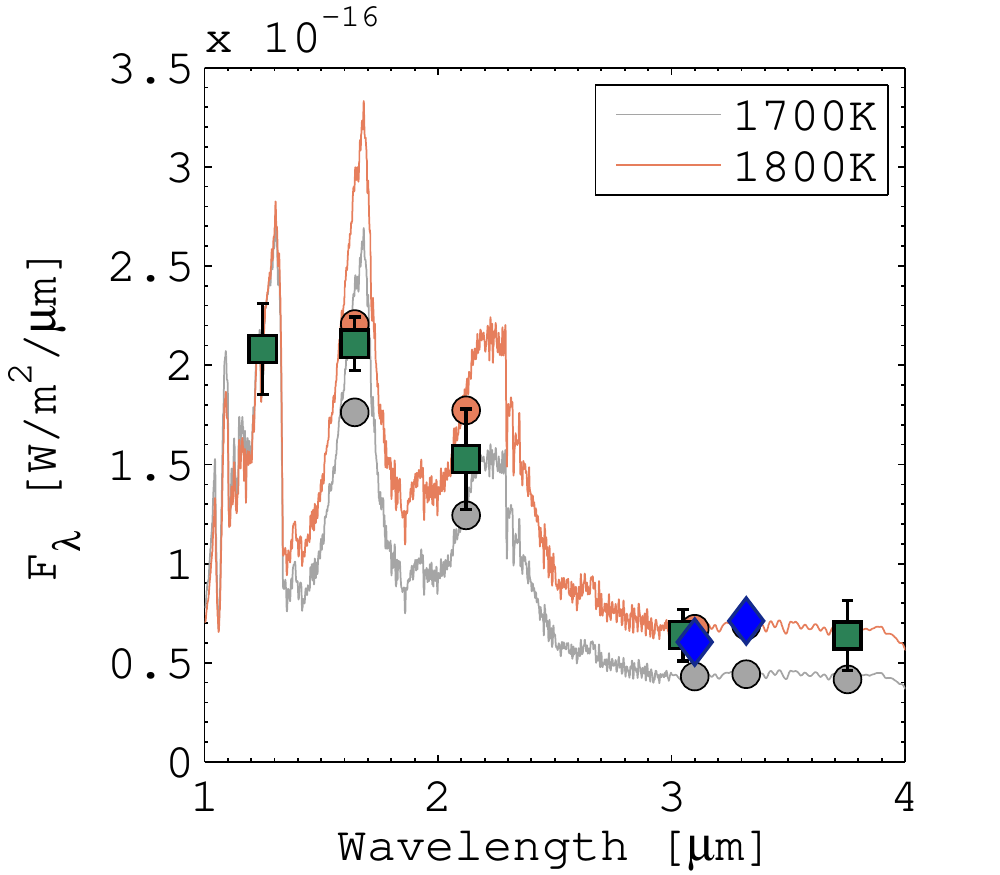}
\caption{1RXS~1609B photometry and DUSTY SED comparison with $A_V=0$. The green squares are the literature photometry and the blue diamonds are this work. Spectra are DUSTY 1800K and 1700K model photospheres with $log(g)=4.0$. The filled circles represent the integrated flux of each spectrum in that bandpass.}
\label{fig:1RXSbSED}
\end{centering}
\end{figure}

Unfortunately, the L2010 sample does not extend past spectral type L0; we therefore compared 1RXS~1609B broadband colors to DUSTY model SEDs with temperatures of $1700K$ and $1800K$.  A surface gravity of $log(g)=4.0$ is predicted by the DUSTY models for these effective temperatures at $5~Myr$. We scaled the SEDs so their integrated $J$ band flux matched the observed flux.  Both SEDs provide an acceptable fit to the data (Figure \ref{fig:1RXSbSED}), consistent with the \citet{Lafreniere2010} effective temperature estimate. Our new photometry confirms the lack of near-IR excess.

\subsubsection{$24\um$ Flux}

Following the procedure used in Section \ref{sect:GSC24umexcess}, we find evidence for a moderate $24\um$ excess in the system. The observed flux is $3.06\pm0.04~mJy$, and the predicted flux from ``A'' is $2.15\pm0.07~mJy$, an excess of $42\pm6\%$ above the expected stellar photospheric emission. The error is dominated by the uncertainty in the literature V magnitude. As with GSC~06214, the predicted contribution to the excess from the BD's photosphere is $<1\%$. Neither ``A'' nor ``B'' exhibits a near-infrared excess, and the components have not been resolved at wavelengths longer than $L^\prime$. Thus, though we can detect the presence of a warm excess in the system, we cannot determine whether it is around the primary, BD, or both.

\subsubsection{Constraints on additional objects}
\label{sect:1RXScompanions}

We do not detect any additional objects in the 1RXS~1609 system. Our background limit at $3.3\um$ was $15.6~mag$, or $\sim5~M_{Jup}$, extending out to $500~AU$. We do not improve upon previous detection limits of $8/1~M_{Jup}$ at $50/450~AU$ stellocentric distance \citep{Lafreniere2010}. We could have detected an equally luminous binary companion to ``B'' (i.e. ``Bb'') at a separation $>25~AU$ from ``B.''

%%%%%%%%%%%%%%%%%%%%%%%%%%
\subsection{HIP 78530}

\subsubsection{Photometry and Astrometry}

\begin{figure}
\begin{centering}
\epsscale{0.9}
%\plotone{fig7.eps}
\plotone{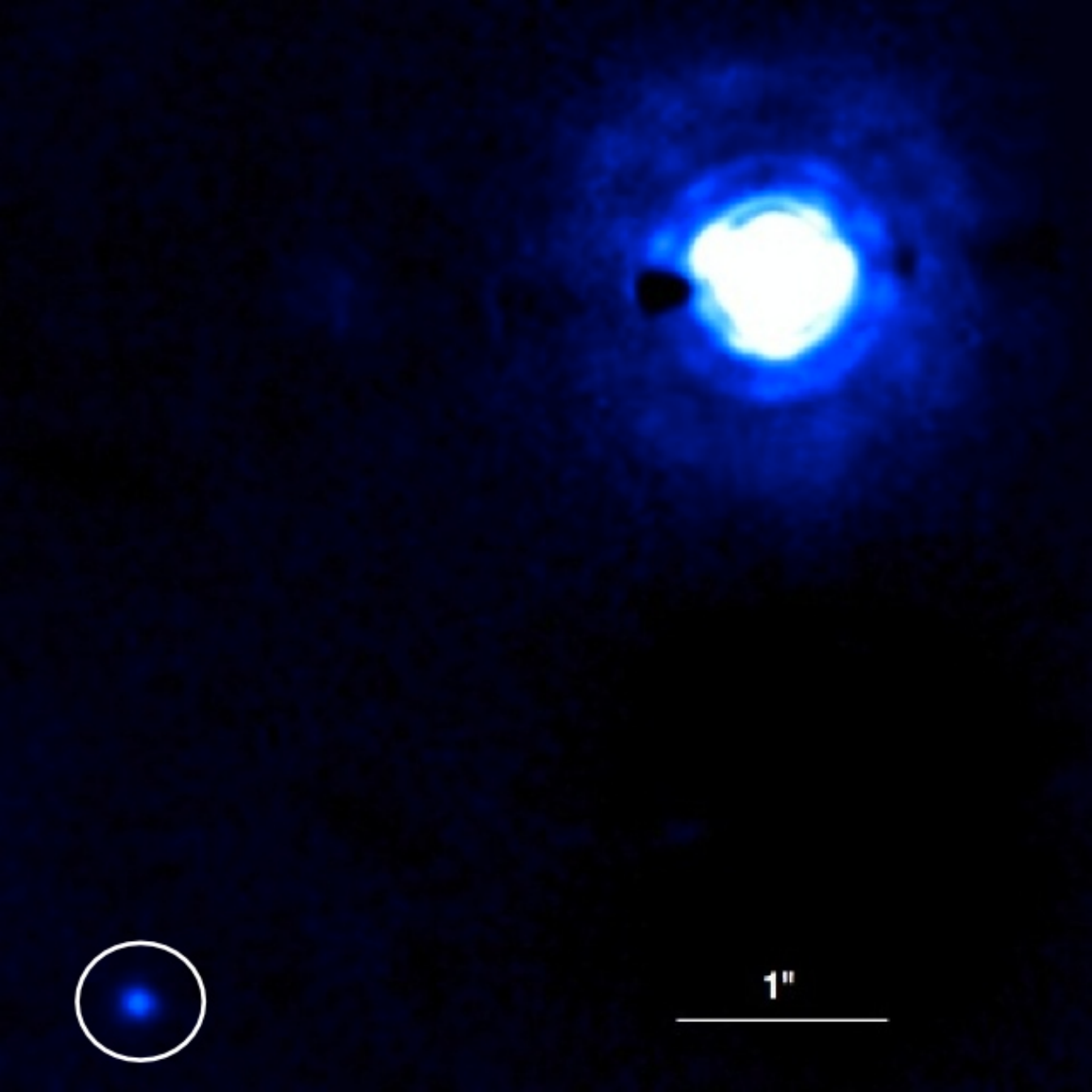}
\caption{HIP~78530 LMIRCam $L^\prime$ image, smoothed for display purposes with a Gaussian kernel with FWHM=$1\lambda/D\sim0.1''$. The companion, circled, is $4.5\arcsec$ to the SE.}
\label{fig:HIP}
\end{centering}
\end{figure}

Figure \ref{fig:HIP} shows the final $L^\prime$ image of HIP~78530, with the BD to the southeast. For the primary, we inferred $L^\prime_A=6.87\pm0.05$. HIP~78530B was detected at a contrast between ``A'' and ``B'' of 6.93 magnitudes (Table \ref{tab:HIP}). The $5\sigma$ sensitivity limit was $L^\prime=16.21$, giving $SNR=47$ and $L^\prime_B= 13.80\pm0.06$. Our measured separation and PA are $4.54\arcsec\pm0.09\arcsec$ and $140.7\dgr\pm1\dgr$, including plate scale and distortion uncertainties (see Section \ref{sect:LBT/LBTIImaging}), consistent with previous observations.

\begin{deluxetable}{lll}
\tablecaption{HIP~78530 system properties and apparent photometry. \label{tab:HIP}}
\tablewidth{0pt}
\tablehead{
\colhead{Property}		& \colhead{HIP~78530A}  & \colhead{HIP~78530B}       	
}
\startdata
Distance $[pc]$ \tablenotemark{a}  		&   \multicolumn{2}{c}{157$\pm$13}  	\\
Spectral type \tablenotemark{b}			& B9V   	  	& M8$\pm$1    	\\
Mass \tablenotemark{b}				&$\sim$2.5 \msun & 23$\pm$3 $M_{Jup}$   \\
$T_{eff}~[K]$   \tablenotemark{b}			& $\sim$10500	& 2800$\pm$200         	\\
$log(L/L_\odot)$  \tablenotemark{b}		& --             	& -2.55$\pm$0.13       	\\
Separation $['']$   \tablenotemark{c}   	&   \multicolumn{2}{c}{4.54$\pm$0.09} 	\\
PA $[\dgr]$ 	 \tablenotemark{c}		&   \multicolumn{2}{c}{140.7$\pm$1}	\\
$J$    	  & 6.928$\pm$0.021 \tablenotemark{d}   & 15.06$\pm$0.05 \tablenotemark{b}    \\
$H$    	  & 6.946$\pm$0.029 \tablenotemark{d}   & 14.39$\pm$0.04 \tablenotemark{b}	\\
$K_s$    	  & 6.903$\pm$0.020 \tablenotemark{d}   & 14.17$\pm$0.04 \tablenotemark{b}	\\
$3.4\um$ \tablenotemark{e}              & \multicolumn{2}{c}{6.842$\pm$0.034}    \\
$L^\prime$ \tablenotemark{c,f}  	& 6.87$\pm$0.05       & 13.80$\pm$0.06               \\
$4.5\um$ \tablenotemark{g}  		& \multicolumn{2}{c}{6.88$\pm$0.01}          \\
$4.6\um$ \tablenotemark{e}              & \multicolumn{2}{c}{6.890$\pm$0.019}    \\
$8.0\um$ \tablenotemark{g}  		& \multicolumn{2}{c}{6.948$\pm$0.008}        \\
$12\um$  \tablenotemark{d}              & \multicolumn{2}{c}{6.924$\pm$0.017}    \\
$24\um$  \tablenotemark{h}  		& \multicolumn{2}{c}{6.846$\pm$0.01}         \\
\enddata

\tablenotetext{a}{Hipparcos catalog \citep{VanLeeuwen2007}.}
\tablenotetext{b}{\citet{Lafreniere2011}.}
\tablenotetext{c}{This work.}
\tablenotetext{d}{2MASS $J/H/K_S$ survey \citep{Skrutskie2006}.}
\tablenotetext{e}{System. WISE survey \citep{Wright2010}.}
\tablenotetext{f}{MKO $L^\prime$.} 
\tablenotetext{g}{System. Spitzer IRAC \citep{Carpenter2006a}.}
\tablenotetext{h}{System. Spitzer MIPS \citep{Carpenter2009}.}

\end{deluxetable}

\subsubsection{Spectral Classification and Near-IR Colors}
\label{sect:HIPAnalysis}

The extinction towards HIP~78530 is $A_V=0.5$ \citep{Carpenter2009}, and we dereddened the photometry using this value. We calculated a $K$ magnitude for the BD of $14.12\pm0.06$ following the prescription described in Section \ref{sect:color-color}.

\begin{figure*}
%\plottwo{fig8a.eps}{fig8b.eps}
\plottwo{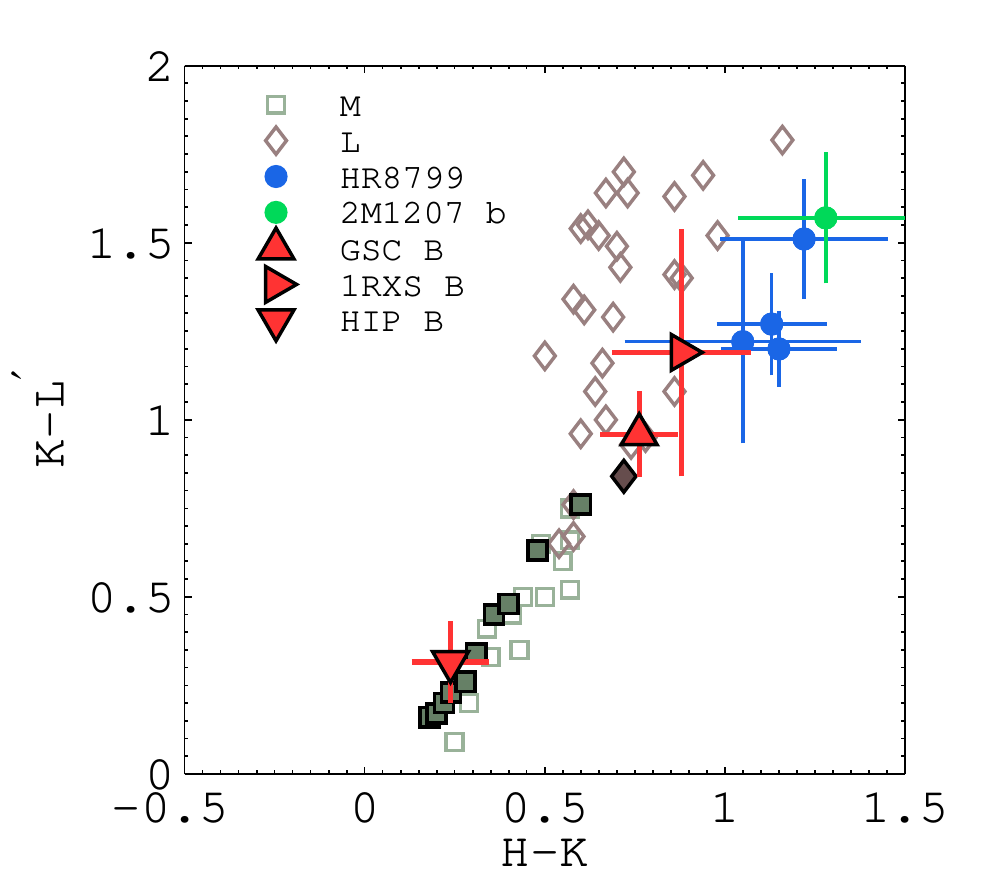}{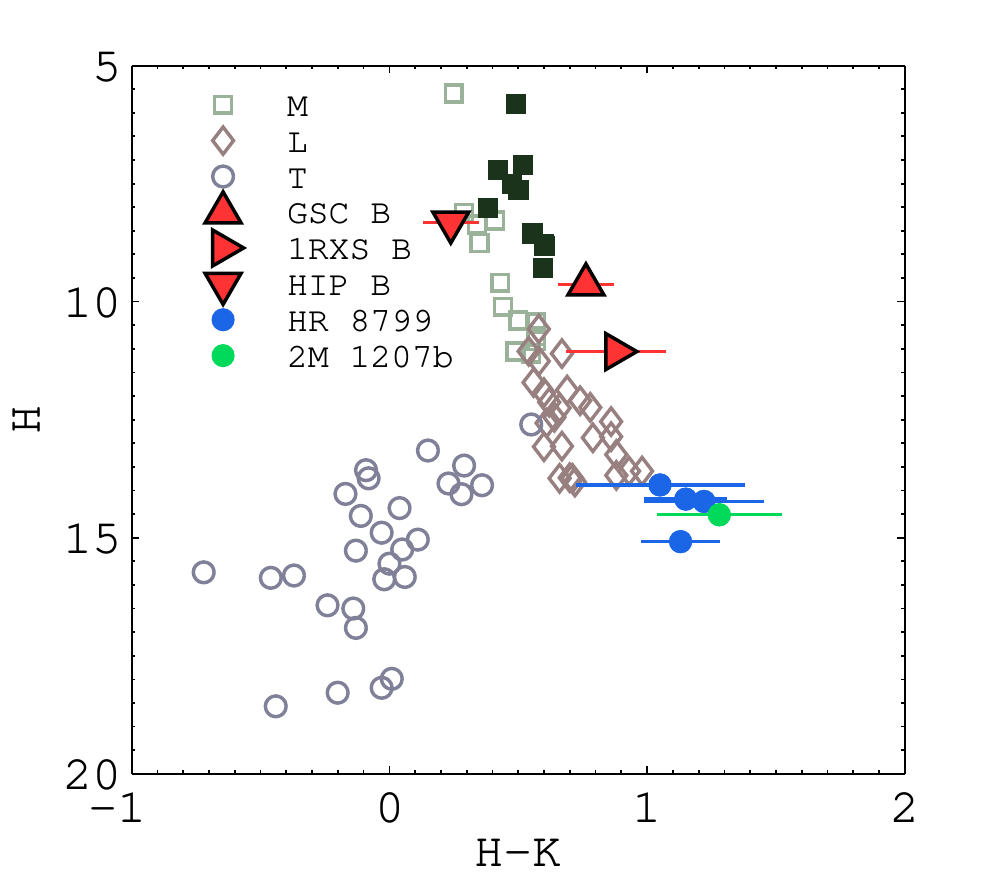}
\caption{$K-L^\prime$ vs.\ $H-K$ color-color diagram \textit{(left)} and $H$ vs.\ $H-K$ color-magnitude diagram \textit{(right)} of dereddened USco companions and comparison samples. Open green squares, brown diamonds, and gray circles: field M, L, and T dwarfs from \protect\citet{Leggett2010}; filled green squares and brown diamond: young M0--L0 dwarfs from L2010; black filled squares: diskless USco objects with spectral between M7--M8.5 \protect\citep{Luhman2012}; filled blue circles: young planets HR~8799bcde \protect\citep{Currie2011b, Marois2008, Marois2010, Skemer2012}; filled green circle: young planet 2M~1207b \protect\citep{Mohanty2007}. }
\label{fig:ccd-cmd}
\end{figure*}

We find the spectral types derived for HIP~78530B from broadband $1-4\um$ colors and from near-IR spectra are not consistent. HIP~78530B was classified using $J/H/K$ spectra as M$8\pm1$ \citep{Lafreniere2011}, though the authors noted the object was $\sim0.2~mag$ too blue at $K$ for this spectral type. We find that its $H-L^\prime$ color is also $\sim0.5~mag$ too blue compared to an L2010 M8-type BD (Figure \ref{fig:colorcolor1}, left panel). There is a similar mismatch in both $H-K$ and $K-L^\prime$ colors (Figure \ref{fig:ccd-cmd}, left panel). With the addition of our new $L^\prime$ data, we find the object's $J$ through $L^\prime$ colors are not consistent with an M8, but instead with an L2010 M$3\pm2$. Intriguingly, the object's absolute magnitude is consistent with other USco M8-type BD \citep{Luhman2012} as shown in Figure \ref{fig:ccd-cmd} (right panel). We discuss the implications of this discrepancy further in Section \ref{sect:HIPDiscussion}. Regardless of whether we adopt spectral class M8 or M3, we find no evidence for $K$ or $L^\prime$ excesses.

\subsubsection{$24\um$ Flux}

The observed and predicted fluxes for the HIP~78530 system are $13.10\pm0.12~mJy$ and $12.0\pm0.5~mJy$. The excess is $9\pm4.5\%$. At $2\sigma$, we do not consider it statistically significant. This, combined with a lack of $1-5\um$ excess around both ``A'' and ``B,'' suggests that neither object retains a massive warm disk.

\subsubsection{Constraints on additional objects}
\label{sect:HIPcompanions}

\begin{figure}
\begin{centering}
%\plotone{fig9.eps}
\plotone{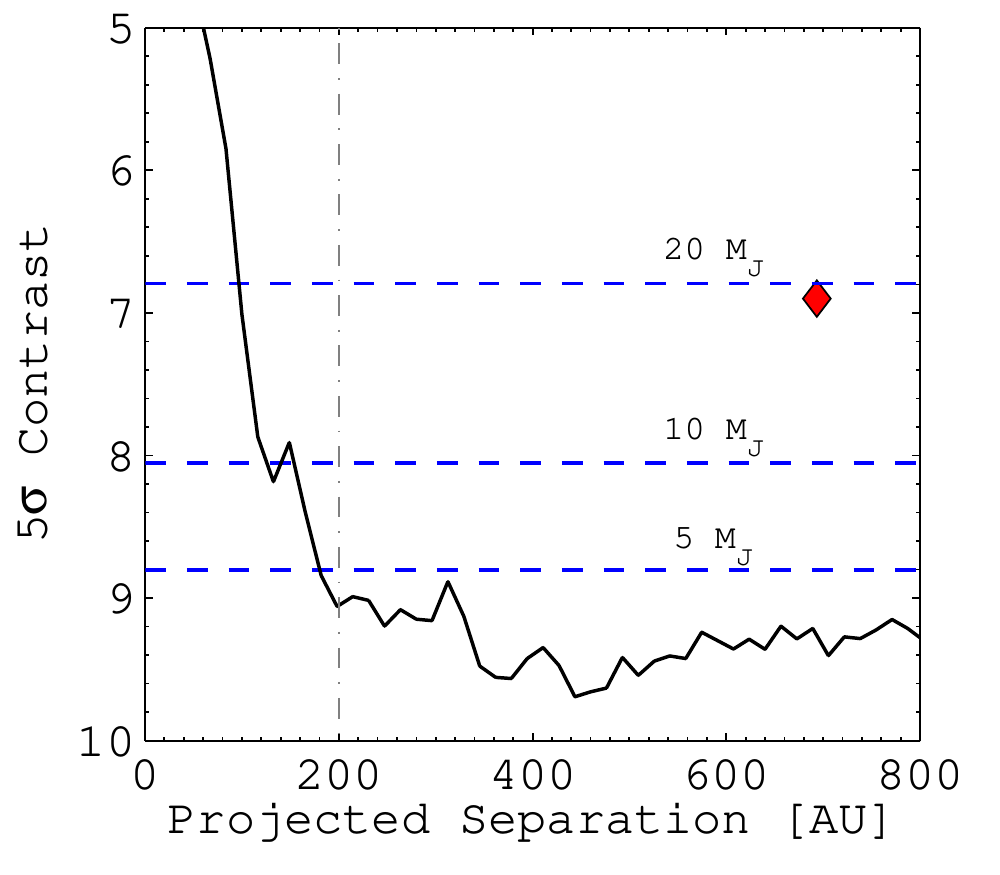}
\caption{LMIRCam $L^\prime~ 5\sigma$ contrast curve for HIP~78530 (black line). Red diamond is HIP~78530B. Dashed lines are apparent magnitudes for objects of various masses according to $5~Myr$ DUSTY models. The dot-dashed line denotes the extent of the FOV on the west half of the system.}
\label{fig:HIPContrast}
\end{centering}
\end{figure}

No additional objects are detected in our data. Figure \ref{fig:HIPContrast} plots our new contrast and mass limits as a function of projected stellocentric distance; we adopt the Hipparcos system distance of $157~pc$ \citep{VanLeeuwen2007}. We reach a sensitivity of $20~M_{Jup}$ beyond $100~AU$ and $\lesssim5~M_{Jup}$ beyond $175~AU$. In the east half of the system, our FOV extended to $800~AU$, while on the west side, the FOV ended at $\sim200~AU$.  An equally luminous binary companion to ``B'' (i.e. ``Bb'') is ruled out at $>15~AU$ separation from ``B.''

%%%%%%%%%%%%%%%%%%%%%%%%%%%%%%%%%%%%%%%%%%%%%%%%%%%%%%%%%

\section{Discussion}
\label{sect:Discussion}

\subsection{GSC~06214B Disk Properties}
\label{sect:GSCBDisk}

GSC~06214B was already known to exhibit an $L^\prime$ excess as well as Pa$\beta$ emission, suggesting a circum(sub)stellar disk \citep{Bowler2011}. We wished to test this hypothesis using the full collection of $1-5\um$ and $24\um$ photometry.  The shape of the $1-5\um$ excess SED is strongly dependent on the temperature (location) of the inner edge of the disk. An accreting dust disk would likely be truncated at the dust sublimation temperature or by magnetic interactions near the corotation radius \citep{Shu1994, Muzerolle2003}, both of which are a few times the BD radius. 

To estimate the location of the disk inner edge, we created simple model rings of blackbody grains at a single temperature. A disk with an inner edge at a temperature of $1550~K$, the sublimation temperature of silicates, reproduced the shape of the observed $J$ through $M^\prime$ excess. Decreasing the temperature to $1000~K$ under-produced $J$, $H$, and $K$ flux relative to the $L^\prime$ and $M^\prime$ flux.  From this we conclude that hot dust must be present within several BD radii away from the BD's surface, consistent with what we expect for an actively accreting disk.

Adding additional weight to the disk hypothesis is the $24\um$ excess. As described in Section \ref{sect:GSC24umexcess}, the GSC~06214 system has an unresolved $1.6~mJy$ excess at $24\um$.   \citet{Luhman2012} recently compiled $K_s-[24]$ and $[4.5]-[24]$ colors for other low-mass USco objects. The ratios of the $24\um$ excess to the $K$ and $M^\prime$ fluxes of GSC~06214B fall within the range that the authors derived for other disk-bearing late M-type objects in the association. Although the primary is shown to have H$\alpha$ emission, presumably due to an accretion disk, the data are consistent with both the near- and mid-IR broadband excesses originating from a single, continuous disk around the BD.

\subsection{HIP~78530B Spectral Type and Membership}
\label{sect:HIPDiscussion}

As discussed in Section \ref{sect:HIPAnalysis}, HIP~78530B's broadband colors do not match the M8 spectral class indicated by the shape of its $J/H/K$ spectra.  The L2010 young photosphere color calibration suggests it is an M$3\pm2$ spectral type (Figures \ref{fig:colorcolor1} and \ref{fig:ccd-cmd}), and the observed colors also more closely match M3V colors than M8V colors (SPEX standards Gl~388 and LP~412-31, respectively). Using the temperature scale calibrated by \citet{Luhman2003}, M3 corresponds to an effective temperature of $3300-3400~K.$ The BCAH $5~Myr$ evolutionary models \citep{Baraffe1998} predict a mass of $0.18-0.4$\msun, depending on whether the $3350~K$ effective temperature, the $H-K$ color, or the $H-L^\prime$ color is used. Regardless of the particular assumptions, these models place the object above the hydrogen burning limit ($\sim75~M_J$).

Conversely, the apparent magnitude of this object is consistent with other M8 brown dwarfs in Upper Sco (Figure \ref{fig:ccd-cmd}) as well as the DUSTY and BCAH predictions for a $5-10~Myr$, $20~M_{Jup}$ BD. The $13~pc$ uncertainty in distance to HIP~78530A cannot account for the faintness of the object compared to other USco M3 BDs. New photometry at $M^\prime$ would help determine whether the object is blue at all wavelengths or merely faint at $K$ and $L^\prime.$

Although the primary has a Hipparcos parallax which is consistent with the distance to USco, it is possible that ``B'' is a background object. In the discovery paper, \citet{Lafreniere2011} found agreement between the strength of gravity-sensitive NIR spectral features in HIP~78530B and those in another USco M8 BD. Additionally, the authors measured common proper motion at the $6\sigma$ level over a 2 year baseline. Unfortunately our astrometric calibration is not sufficiently precise to confirm or deny common proper motion. An additional epoch of observation would help to cement this object's status as a bound companion.

%%%%%%%%%%%%%%%%%%%%%%%%%%%%%%%%%%%%%%%%%%%%%%%%%%%%%%%%

\section{Disks as Constraints on Scattering Events}
\label{sect:DiskHistories}

The presence of an accreting disk around a low-mass companion at a wide separation from its primary star has the potential to limit the possible formation histories of the companion.  To see why this is the case, we consider the fate of a gas disk orbiting a brown dwarf of mass $M$, which is in turn orbiting a star of mass $M_*$.  An encounter or series of encounters with a perturber of mass $M_p > M$, also orbiting the primary star, can scatter the brown dwarf to wide separations, while typically leaving the more massive perturber on an orbit near the primary.  

Each encounter truncates the disk at a distance, $r$, from the brown dwarf.  This truncation distance is roughly the location where the tidal force during the encounter generates a velocity kick $\Delta v \sim (GM_p/b^2)(r/b)(2b/v_{\rm enc})$ equal to the escape velocity from the brown dwarf, $v_{\rm esc} = (2GM/r)^{1/2}$, where $G$ is the gravitational constant, and $b$ and $v_{\rm enc}$ are the distance of closest approach and the relative velocity of the two bodies during the encounter, respectively.  We consider a fiducial encounter for which $b$ is of order the Hill radius of the perturber, $b \sim R_{H,p} = a(M_p/3M_*)^{1/3}$ and $v_{\rm enc} = \Omega b$, is the relative velocity of two circular Keplerian orbits with semi-major axes that differ by $b$.  Here, $a$ is the distance from the host star at which the encounter occurs, and $\Omega = (GM_*/a^3)^{1/2}$ is the angular velocity of a circular orbit at that distance.  These choices are intended to maximize the expected disk truncation during scattering, without invoking unusually close or slow encounters, which we expect to be rare.  Note that arbitrarily extreme encounters are disallowed because the BD cannot be ejected from the system.  Setting $\Delta v = v_{\rm esc}$ yields a truncation radius of $r \sim 0.5 R_H$, where $R_H$ is the brown dwarf's Hill radius. The disk truncation radius from scattering is thus predicted to be comparable to the radius generated merely by long-term tidal interactions with the host star, $\sim 0.4 R_H$ \citep[e.g.][]{Martin2011}. To solidify interpretation of future observations, we suggest that numerical simulations of scattering histories producing wide-separation BD companions should be mined to determine the typical properties of their most disruptive close encounters.

Although scattering does not significantly strip disk material from the BD, the disk can still be used to constrain the BD's scattering history. This is because the disk size, $\sim 0.4 R_H$, is directly proportional to $a$, the semi-major axis at which the BD formed. We next estimate the range of formation locations which would produce disks large enough to survive for $5~Myr$, the minimum age of USco.

Observations of protoplanetary disks at separations of tens of AU yield surface density profiles of order $\Sigma = 10^3$~g/cm$^2$~$(r/{\rm AU})^{-1}$ \citep{Andrews2009}, comparable to the minimum-mass solar nebula.  Disks around brown dwarfs are likely less massive, so this profile places an upper limit on the disk lifetime.  If such profiles extend inward to smaller separations from the BD, a typical disk contains a mass of $M_d \sim 7\times10^{-4} M_\odot (r/{\rm AU}) $ within a radius $r$. The size of the inner clearing of an accreting disk is negligible, as we found in Section \ref{sect:GSCBDisk}.  Assuming that a typical low-mass brown dwarf has an accretion rate comparable to GSC~06214B, $\dot M = 2\times 10^{-11} M_\odot/yr$ \citep{Bowler2011}, it would accrete at least $10^{-4} M_\odot$ of material over $5~Myr$, the estimated mass interior to $r = 0.15~AU$.  A higher accretion rate of $10^{-10} M_\odot/yr$ would process $5\times10^{-4} M_\odot$ over $5~Myr$, equal to the estimated mass within $r=0.7~AU$.  

These estimates of the minimum disk size can be used to constrain the formation location of the BD. A truncation radius of $0.4a(M/3M_*)^{1/3}$ corresponds to $0.06a-0.07a$ for our three BDs, using the masses listed in Tables \ref{tab:GSC}, \ref{tab:1RXS}, and \ref{tab:HIP}. We estimate that insufficient mass would remain in the current disk orbiting GSC~06214B if the brown dwarf were scattered outward by a perturber or perturbers from a stellocentric distance of $a \lesssim 2-10~AU$.  Current limits on additional companions in the GSC~06214 system can rule out $35M_{Jup} \sim 2M_B$ objects beyond $10~AU$ stellocentric distance (Section \ref{sect:GSCcompanions}), disfavoring, though note entirely ruling out, the scattering hypothesis. Limits on the minimum semi-major axis at formation of HIP~78530B and 1RXS~1609B, based on their lack of accretion signatures, are similar, presuming they had accretion rates similar to that of GSC~06214B. Additional objects more than twice as massive as the known BDs are ruled out beyond 35AU for 1RXS~1609 and beyond $75~AU$ for HIP~78530 (\citet{Lafreniere2010} and \ref{sect:HIPcompanions}). These weaker constraints cannot be used to rule out a scattering origin for either object.  Given our estimates, tighter observational constraints on the presence of possible perturbers,  coupled with more detailed modeling of disk truncation during ejection by scattering will provide significant limits on the scattering histories of wide-separation, low-mass companions.

%%%%%%%%%%%%%%%%%%%%%%%%%%%%%%%%%%%%%%%%%%%%%%%%%%%%%%%%

\section{Summary and Future Work}
\label{Summary}

We present a  3--5 $\mu m$ LBT/MMT imaging study of three low-mass (ratio) companions to Upper Scorpius stars: GSC~06214B, 1RXS~1609B, and HIP~78530B. These three systems constitute some of the lowest mass-ratio ($q\sim1\%$) non-planetary systems discovered to date. However, though the companions have similar separations and mass ratios, they probe two very different types of binary systems: GSC~06214 and 1RXS~1609 are K-type/L-type pairs, while HIP~78530 is a B-type/M-type pair. Combined with MIPS-24 $\mu m$ photometry, we use our new high-resolution near/thermal IR data to identify evidence for broadband disk emission from small dust grains around both the companions and the primary stars.  Our study yields the following major results:
\begin{itemize}

\item We confirm the presence of a circum(sub)stellar disk around the L0-type BD GSC~06214B. Furthermore, we find the best fit temperature of the disk inner edge is at the dust sublimation temperature, consistent with active accretion. A moderate $24\um$ excess in the system is plausibly explained by the contribution from the BD's disk, although some material must still be present around the primary in order to explain its H$\alpha$ emission.  The existence of a disk around GSC~06214B suggests, but does not conclusively prove, it has not undergone a scattering event.
 
 \item We find evidence of a weak $24\um$ excess in the 1RXS~1609 system, but the $<4\um$ photometry of the primary and low-mass companion are consistent with purely photospheric emission. The $24\um$ excess is unresolved, so we cannot determine whether ``A,'' ``B,'' or both host a warm disk. 
  
\item In contrast, we find no evidence of $1-4\um$ or $24\um$ excesses in the HIP~78530 system. However, we find a discrepancy between the spectral types indicated by the $1-4\um$ broadband colors and by the $J/H/K$ spectra. Although the spectra and apparent magnitude are consistent with an USco M8-type BD, the broadband colors instead match a young M3-type low-mass star. Future work corroborating common proper motion in the system is necessary to decisively rule out the possibility that ``B'' is a background object.  We present improved constraints on the presence of additional low-mass companions, ruling out equally luminous companions at projected separations from the primary $>100~AU$ and reaching a background limit of $<5~M_{Jup}$ beyond $175~AU$.

\end{itemize}

Although our sample size is too small for robust statistics, the detection of a massive, hot disk in one of three low-mass companions is consistent with the disk fraction determined for single BDs in USco \citep{Riaz2012, Scholz2007, Luhman2012}.   Furthermore, though the range and relative importance of formation mechanisms for low-mass, widely separated companions remains unclear, this work does provide some constraints on the formation of our USco targets.   

In particular, we corroborate the findings of \citet{Bowler2011}, who identified a disk around GSC~06214B from accretion-driven Pa$\beta$ emission. 
Our study further confirms that this disk is dusty, producing a strong broadband IR excess from grains near the dust sublimation radius, much like for many disks around stars in USco \citep{Carpenter2006a,Carpenter2009, Luhman2012}.  The presence of a dusty, accreting disk around the BD, combined with the constraint that no object more than twice as massive as ``B'' exists beyond $10~AU$ stellocentric distance, casts doubt on whether this companion could have been formed via a planet-planet scattering event \citep[i.e. as in][]{Veras2009}.   

We do not detect a near-IR broadband excess around 1RXS~1609B. (The fact that the L4-type 1RXS~1609B does not have a hot disk, while the more massive L0-type GSC~06214B does, need not be contradictory; it is within the scatter in the inner disk dissipation timescale.) We do, however, find the system as a whole exhibits a $24~\um$ excess, presumably from warm dust.  If some of this emission originates from a large disk surrounding the BD companion, then a scattering origin for this BD may also be disfavored. 

Conversely, a lack of $24\um$ excess around the more massive HIP~78530B does not necessarily imply a past scattering event, as depletion of warm dust from passive disk evolution alone is more rapid for intermediate-mass than for low-mass BDs \citep{Luhman2012}.

Resolved observations with ALMA and/or NOMIC (the $8-13\um$ channel of LBTI) would be able to further constrain the disk properties of both the GSC~06214 primary and BD and could clarify whether the 1RXS~1609 system's warm $24\um$ excess emission originates solely from  material around the primary or around the low-mass companion as well. ALMA observations could in principle identify even colder dust  emission from outer disk regions and provide a direct measurement of disk mass. They might also be used to differentiate between fully depleted disks and those with central clearings. A comparison of the incidence, masses, and morphologies derived in this way for disks around a larger sample of single and companion BDs might discriminate between in situ formation and scattering events.  Finally, comparing cluster mass functions with age may identify a time-dependent low-mass BD population consistent with some fraction of low-mass BDs having a scattering origin.

\acknowledgments
 We gratefully acknowledge the hard work and dedication of all members of the LBTI team including: Tom McMahon, Paul Arbo, Teresa Bippert-Plymate, Oli Durney, Manny Montoya, Mitch Nash, and Elliot Solheid. We thank the anonymous referee for their helpful comments; Daniel Apai, Laird Close, Michael Cushing, and Kevin Luhman for their insightful discussions; Brendan Bowler for sharing his GSC~06214A SPEX data; and the LBT and MMT staff for their support of these observations. This work is based in part on observations made with the Spitzer Space Telescope, which is operated by the Jet Propulsion Laboratory, California Institute of Technology under a contract with NASA. We utilized data products from the Wide-field Infrared Survey Explorer, which is a joint project of the University of California, Los Angeles, and the Jet Propulsion Laboratory/California Institute of Technology, funded by the National Aeronautics and Space Administration. We also utilized data products from the Two Micron All Sky Survey, which is a joint project of the University of Massachusetts and the Infrared Processing and Analysis Center/California Institute of Technology, funded by the National Aeronautics and Space Administration and the National Science Foundation. LBTI is funded by a NASA grant in support of the Exoplanet Exploration Program (NSF 0705296). V.B. is supported by the National Science Foundation Graduate Research Fellowship (NSF DGE-1143953).

\bibliography{bailey}

\begin{thebibliography}{85}
\expandafter\ifx\csname natexlab\endcsname\relax\def\natexlab#1{#1}\fi

\bibitem[{Andrews {et~al.}(2009)Andrews, Wilner, Hughes, Qi, \&
  Dullemond}]{Andrews2009}
Andrews, S.~M., Wilner, D.~J., Hughes, a.~M., Qi, C., \& Dullemond, C.~P. 2009,
  The Astrophysical Journal, 700, 1502

\bibitem[{Baraffe {et~al.}(1998)Baraffe, Chabrier, Allard, \&
  Hauschildt}]{Baraffe1998}
Baraffe, I., Chabrier, G., Allard, F., \& Hauschildt, P.~H. 1998, Astronomy \&
  Astrophysics, 337, 403

\bibitem[{Barman {et~al.}(2011)Barman, Macintosh, Konopacky, \&
  Marois}]{Barman2011}
Barman, T., Macintosh, B., Konopacky, Q., \& Marois, C. 2011, The Astrophysical
  Journal Letters, 735, L39

\bibitem[{Bate(2000)}]{Bate2000}
Bate, M.~R. 2000, Monthly Notices of the Royal Astronomical Society, 314, 33

\bibitem[{Boss(2011)}]{Boss2011}
Boss, A.~P. 2011, The Astrophysical Journal, 731, 74

\bibitem[{Bowler {et~al.}(2011)Bowler, Liu, Kraus, Mann, \&
  Ireland}]{Bowler2011}
Bowler, B.~P., Liu, M.~C., Kraus, A.~L., Mann, A.~W., \& Ireland, M.~J. 2011,
  The Astrophysical Journal, 743, 148

\bibitem[{Brusa {et~al.}(2004)Brusa, Miller, Kenworthy, Fisher, \&
  Riccardi}]{Brusa2004}
Brusa, G., Miller, D.~L., Kenworthy, M., Fisher, D., \& Riccardi, A. 2004,
  Proceedings of SPIE, 5490, 23

\bibitem[{Burgasser {et~al.}(2003)Burgasser, Kirkpatrick, Reid, Brown, Miskey,
  \& Gizis}]{Burgasser2003}
Burgasser, A.~J., Kirkpatrick, J.~D., Reid, I.~N., Brown, M.~E., Miskey, C.~L.,
  \& Gizis, J.~E. 2003, The Astrophysical Journal, 586, 512

\bibitem[{Canup \& Ward(2002)}]{canup2002}
Canup, R., \& Ward, W. 2002, The Astronomical Journal, 124, 3404

\bibitem[{Cardelli {et~al.}(1989)Cardelli, Clayton, \& Mathis}]{Cardelli1989}
Cardelli, J.~A., Clayton, G.~C., \& Mathis, J.~S. 1989, The Astrophysical
  Journal, 345, 245

\bibitem[{Carpenter {et~al.}(2006)Carpenter, Mamajek, Hillenbrand, \&
  Meyer}]{Carpenter2006a}
Carpenter, J.~M., Mamajek, E.~E., Hillenbrand, L.~A., \& Meyer, M.~R. 2006, The
  Astrophysical Journal Letters, 651, L49

\bibitem[{Carpenter {et~al.}(2009)Carpenter, Mamajek, Hillenbrand, \&
  Meyer}]{Carpenter2009}
Carpenter, J.~M., Mamajek, E.~E., Hillenbrand, L.~a., \& Meyer, M.~R. 2009, The
  Astrophysical Journal, 705, 1646

\bibitem[{Carson {et~al.}(2013)Carson, Thalmann, Janson, Kozakis, Bonnefoy,
  Biller, Schlieder, Currie, McElwain, Goto, Henning, Brandner, Feldt, Kandori,
  Kuzuhara, Stevens, Wong, Gainey, Fukagawa, Kuwada, Brandt, Kwon, Abe, Egner,
  Grady, Guyon, Hashimoto, Hayano, Hayashi, Hayashi, Hodapp, Ishii, Iye, Knapp,
  Kudo, Kusakabe, Matsuo, Miyama, Morino, Moro-Martin, Nishimura, Pyo, Serabyn,
  Suto, Suzuki, Takami, Takato, Terada, Turner, Watanabe, Wisniewski, Yamada,
  Takami, Usuda, \& Tamura}]{Carson2013}
Carson, J., {et~al.} 2013, The Astrophysical Journal Letters, 763, L32

\bibitem[{Chabrier {et~al.}(2000)Chabrier, Baraffe, Allard, \&
  Hauschildt}]{Chabrier2000}
Chabrier, G., Baraffe, I., Allard, F., \& Hauschildt, P. 2000, The
  Astrophysical Journal, 542, 464

\bibitem[{Chauvin {et~al.}(2004)Chauvin, Lagrange, Dumas, Zuckerman, Mouillet,
  Song, Beuzit, \& Lowrance}]{Chauvin2004}
Chauvin, G., Lagrange, A., Dumas, C., Zuckerman, B., Mouillet, D., Song, I.,
  Beuzit, J., \& Lowrance, P. 2004, Astronomy \& Astrophysics, 425, L29

\bibitem[{Close {et~al.}(2003)Close, Siegler, Freed, \& Biller}]{Close2003}
Close, L.~M., Siegler, N., Freed, M., \& Biller, B. 2003, The Astrophysical
  Journal, 587, 407

\bibitem[{Currie {et~al.}(2009)Currie, Lada, Plavchan, Robitaille, Irwin, \&
  Kenyon}]{Currie2009}
Currie, T., Lada, C.~J., Plavchan, P., Robitaille, T.~P., Irwin, J., \& Kenyon,
  S.~J. 2009, The Astrophysical Journal, 698, 1

\bibitem[{Currie {et~al.}(2011)Currie, Burrows, Itoh, Matsumura, Fukagawa,
  Apai, Madhusudhan, Hinz, Rodigas, Kasper, Pyo, \& Ogino}]{Currie2011b}
Currie, T., {et~al.} 2011, The Astrophysical Journal, 729, 128

\bibitem[{Currie {et~al.}(2012)Currie, Debes, Rodigas, Burrows, Itoh, Fukagawa,
  Kenyon, Kuchner, \& Matsumura}]{Currie2012}
---. 2012, The Astrophysical Journal, 760, L32

\bibitem[{Dawson {et~al.}(2013)Dawson, Scholz, Ray, Marsh, Wood, Natta,
  Padgett, \& Ressler}]{Dawson2013}
Dawson, P., Scholz, A., Ray, T.~P., Marsh, K.~a., Wood, K., Natta, A., Padgett,
  D., \& Ressler, M.~E. 2013, Monthly Notices of the Royal Astronomical
  Society, 429, 903

\bibitem[{de~Zeeuw {et~al.}(1999)de~Zeeuw, Hoogerwerf, de~Bruijne, Brown, \&
  Blaauw}]{DeZeeuw1999}
de~Zeeuw, P.~T., Hoogerwerf, R., de~Bruijne, J. H.~J., Brown, A. G.~A., \&
  Blaauw, A. 1999, The Astronomical Journal, 117, 354

\bibitem[{Diolaiti {et~al.}(2000)Diolaiti, Bendinelli, Bonaccini, Close,
  Currie, \& Parmeggiani}]{Diolaiti2000}
Diolaiti, E., Bendinelli, O., Bonaccini, D., Close, L.~M., Currie, D.~G., \&
  Parmeggiani, G. 2000, Proceedings of SPIE, 4007, 879

\bibitem[{Engelbracht {et~al.}(2007)Engelbracht, Blaylock, Su, Rho, Rieke,
  Muzerolle, Padgett, Hines, Gordon, Fadda, Noriega-Crespo, Kelly, Latter,
  Hinz, Misselt, Morrison, Stansberry, Shupe, Stolovy, Wheaton, Young,
  Neugebauer, Wachter, Perez-Gonzalez, Frayer, \& Marleau}]{Engelbracht2007}
Engelbracht, C.~W., {et~al.} 2007, Publications of the Astronomical Society of
  the Pacific, 119, 994

\bibitem[{Esposito {et~al.}(2011)Esposito, Riccardi, Pinna, Puglisi,
  Quiros-Pacheco, Arcidiacono, Xompero, Briguglio, Agapito, Busoni, Fini,
  Argomedo, Gherardi, Brusa, Miller, Guerra, Stefanini, \&
  Salinari}]{Esposito2011}
Esposito, S., {et~al.} 2011, Proceedings of SPIE, 8149, 814902

\bibitem[{Fischer \& Marcy(1992)}]{Fischer1992}
Fischer, D.~A., \& Marcy, G.~W. 1992, The Astrophysical Journal, 396, 178

\bibitem[{Galicher {et~al.}(2011)Galicher, Marois, Macintosh, Barman, \&
  Konopacky}]{Galicher2011}
Galicher, R., Marois, C., Macintosh, B., Barman, T., \& Konopacky, Q. 2011, The
  Astrophysical Journal, 739, L41

\bibitem[{Gordon {et~al.}(2005)Gordon, Rieke, Engelbracht, Muzerolle,
  Stansberry, Misselt, Morrison, Cadien, Young, Dole, Kelly, Alonso-Herrero,
  Egami, Su, Papovich, Smith, Hines, Rieke, Blaylock, Perez-Gonzalez, {Le
  Floc'h}, Hinz, Latter, Hesselroth, Frayer, Noriega-Crespo, Masci, Padgett,
  Smylie, \& Haegel}]{Gordon2005}
Gordon, K., {et~al.} 2005, Publications of the Astronomical Society of the
  Pacific, 117, 503

\bibitem[{Hinz {et~al.}(2008)Hinz, Solheid, Durney, \& Hoffmann}]{Hinz2008}
Hinz, P.~M., Solheid, E., Durney, O., \& Hoffmann, W.~F. 2008, Proceedings of
  SPIE, 7013, 701339

\bibitem[{Ireland {et~al.}(2011)Ireland, Kraus, Martinache, Law, \&
  Hillenbrand}]{Ireland2011}
Ireland, M.~J., Kraus, A., Martinache, F., Law, N., \& Hillenbrand, L. 2011,
  The Astrophysical Journal, 726, 113

\bibitem[{Janson {et~al.}(2012{\natexlab{a}})Janson, Jayawardhana, Girard,
  Lafreni\`{e}re, Bonavita, Gizis, \& Brandeker}]{Janson2012a}
Janson, M., Jayawardhana, R., Girard, J.~H., Lafreni\`{e}re, D., Bonavita, M.,
  Gizis, J., \& Brandeker, A. 2012{\natexlab{a}}, The Astrophysical Journal,
  758, L2

\bibitem[{Janson {et~al.}(2012{\natexlab{b}})Janson, Hormuth, Bergfors,
  Brandner, Hippler, Daemgen, Kudryavtseva, Schmalzl, Schnupp, \&
  Henning}]{Janson2012b}
Janson, M., {et~al.} 2012{\natexlab{b}}, The Astrophysical Journal, 754, 44

\bibitem[{Jensen {et~al.}(1996)Jensen, Mathieu, \& Fuller}]{Jensen1996}
Jensen, E. L.~N., Mathieu, R.~D., \& Fuller, G.~A. 1996, The Astrophysical
  Journal, 458, 312

\bibitem[{Joergens(2008)}]{Joergens2008}
Joergens, V. 2008, Astronomy \& Astrophysics, 492, 545

\bibitem[{Kalas {et~al.}(2008)Kalas, Graham, Chiang, Fitzgerald, Clampin, Kite,
  Stapelfeldt, Marois, \& Krist}]{Kalas2008}
Kalas, P., {et~al.} 2008, Science, 322, 1345

\bibitem[{Kratter {et~al.}(2010)Kratter, Murray-Clay, \& Youdin}]{Kratter2010}
Kratter, K.~M., Murray-Clay, R.~a., \& Youdin, A.~N. 2010, The Astrophysical
  Journal, 710, 1375

\bibitem[{Kraus {et~al.}(2012)Kraus, Ireland, Hillenbrand, \&
  Martinache}]{Kraus2012}
Kraus, A., Ireland, M., Hillenbrand, L., \& Martinache, F. 2012, The
  Astrophysical Journal, 745, 19

\bibitem[{Kraus {et~al.}(2008)Kraus, Ireland, Martinache, \& Lloyd}]{Kraus2008}
Kraus, A.~L., Ireland, M.~J., Martinache, F., \& Lloyd, J.~P. 2008, The
  Astrophysical Journal, 679, 762

\bibitem[{Lafreni\`{e}re {et~al.}(2011)Lafreni\`{e}re, Jayawardhana, Janson,
  Helling, Witte, \& Hauschildt}]{Lafreniere2011}
Lafreni\`{e}re, D., Jayawardhana, R., Janson, M., Helling, C., Witte, S., \&
  Hauschildt, P. 2011, The Astrophysical Journal, 730, 42

\bibitem[{Lafreni\`{e}re {et~al.}(2008)Lafreni\`{e}re, Jayawardhana, \& {Van
  Kerkwijk}}]{Lafreniere2008}
Lafreni\`{e}re, D., Jayawardhana, R., \& {Van Kerkwijk}, M. 2008, The
  Astrophysical Journal Letters, 689, L153

\bibitem[{Lafreni\`{e}re {et~al.}(2010)Lafreni\`{e}re, Jayawardhana, \& {Van
  Kerkwijk}}]{Lafreniere2010}
Lafreni\`{e}re, D., Jayawardhana, R., \& {Van Kerkwijk}, M.~H. 2010, The
  Astrophysical Journal, 719, 497

\bibitem[{Lagrange {et~al.}(2010)Lagrange, Bonnefoy, Chauvin, Apai, Ehrenreich,
  Boccaletti, Gratadour, Rouan, Mouillet, Lacour, \& Kasper}]{Lagrange2010}
Lagrange, A.-M., {et~al.} 2010, Science, 329, 57

\bibitem[{Leggett {et~al.}(2010)Leggett, Burningham, \& Saumon}]{Leggett2010}
Leggett, S.~K., Burningham, B., \& Saumon, D. 2010, The Astrophysical Journal,
  710, 1627

\bibitem[{Leggett {et~al.}(2006)Leggett, Currie, Varricatt, Hawarden, Adamson,
  Buckle, Carroll, Davies, Davis, Kerr, Kuhn, Seigar, \& Wold}]{Leggett2006}
Leggett, S.~K., {et~al.} 2006, Monthly Notices of the Royal Astronomical
  Society, 373, 781

\bibitem[{Lloyd-Hart(2000)}]{Lloyd-Hart2000}
Lloyd-Hart, M. 2000, Publications of the Astronomical Society of the Pacific,
  112, 264

\bibitem[{Luhman {et~al.}(2010)Luhman, Allen, Espaillat, Hartmann, \&
  Calvet}]{Luhman2010}
Luhman, K.~L., Allen, P.~R., Espaillat, C., Hartmann, L., \& Calvet, N. 2010,
  The Astrophysical Journal Supplement Series, 186, 111

\bibitem[{Luhman {et~al.}(2008)Luhman, Hern\'{a}ndez, Downes, Hartmann, \&
  Brice\~{n}o}]{Luhman2008}
Luhman, K.~L., Hern\'{a}ndez, J., Downes, J.~J., Hartmann, L., \& Brice\~{n}o,
  C. 2008, The Astrophysical Journal, 688, 362

\bibitem[{Luhman \& Mamajek(2012)}]{Luhman2012}
Luhman, K.~L., \& Mamajek, E.~E. 2012, The Astrophysical Journal, 758, 31

\bibitem[{Luhman {et~al.}(2003)Luhman, Stauffer, Muench, Rieke, Lada, Bouvier,
  \& Lada}]{Luhman2003}
Luhman, K.~L., Stauffer, J.~R., Muench, A.~A., Rieke, G.~H., Lada, E.~A.,
  Bouvier, J., \& Lada, C.~J. 2003, The Astrophysical Journal, 593, 1093

\bibitem[{Luhman {et~al.}(2006)Luhman, Wilson, Brandner, Skrutskie, Nelson,
  Smith, Peterson, Cushing, \& Young}]{Luhman2006}
Luhman, K.~L., {et~al.} 2006, The Astrophysical Journal, 649, 894

\bibitem[{Lunine \& Stevenson(1982)}]{Lunine1982}
Lunine, J.~I., \& Stevenson, D.~J. 1982, Icarus, 52, 14

\bibitem[{Marcy \& Butler(2000)}]{Marcy2000}
Marcy, G.~W., \& Butler, R.~P. 2000, Publications of the Astronomical Society
  of the Pacific, 112, 137

\bibitem[{Marois {et~al.}(2006)Marois, Lafreni\`{e}re, Doyon, Macintosh, \&
  Nadeau}]{Marois2006}
Marois, C., Lafreni\`{e}re, D., Doyon, R., Macintosh, B., \& Nadeau, D. 2006,
  The Astrophysical Journal, 641, 556

\bibitem[{Marois {et~al.}(2008{\natexlab{a}})Marois, Lafreniere, Macintosh, \&
  Doyon}]{Marois2008}
Marois, C., Lafreniere, D., Macintosh, B., \& Doyon, R. 2008{\natexlab{a}}, The
  Astrophysical Journal, 673, 647

\bibitem[{Marois {et~al.}(2008{\natexlab{b}})Marois, Macintosh, Barman,
  Zuckerman, Song, Patience, Lafreni\`{e}re, \& Doyon}]{Marois2008a}
Marois, C., Macintosh, B., Barman, T., Zuckerman, B., Song, I., Patience, J.,
  Lafreni\`{e}re, D., \& Doyon, R. 2008{\natexlab{b}}, Science, 322, 1348

\bibitem[{Marois {et~al.}(2010)Marois, Zuckerman, Konopacky, Macintosh, \&
  Barman}]{Marois2010}
Marois, C., Zuckerman, B., Konopacky, Q.~M., Macintosh, B., \& Barman, T. 2010,
  Nature, 468, 1080

\bibitem[{Martin \& Lubow(2011)}]{Martin2011}
Martin, R.~G., \& Lubow, S.~H. 2011, Monthly Notices of the Royal Astronomical
  Society, 413, 1447

\bibitem[{Metchev \& Hillenbrand(2006)}]{Metchev2006}
Metchev, S.~A., \& Hillenbrand, L.~A. 2006, The Astrophysical Journal, 651,
  1166

\bibitem[{Mohanty {et~al.}(2007)Mohanty, Jayawardhana, Huelamo, \&
  Mamajek}]{Mohanty2007}
Mohanty, S., Jayawardhana, R., Huelamo, N., \& Mamajek, E.~E. 2007, The
  Astrophysical Journal, 657, 1064

\bibitem[{Monin {et~al.}(2010)Monin, Guieu, Pinte, Rebull, Goldsmith, Fukagawa,
  M\'{e}nard, Padgett, Stappelfeld, McCabe, Carey, Noriega-Crespo, Brooke,
  Huard, Terebey, Hillenbrand, \& Guedel}]{Monin2010}
Monin, J.-L., {et~al.} 2010, Astronomy and Astrophysics, 515, A91

\bibitem[{Muzerolle {et~al.}(2003)Muzerolle, Hillenbrand, Calvet, Brice\~{n}o,
  \& Hartmann}]{Muzerolle2003}
Muzerolle, J., Hillenbrand, L.~A., Calvet, N., Brice\~{n}o, C., \& Hartmann, L.
  2003, The Astrophysical Journal, 592, 266

\bibitem[{Pascucci {et~al.}(2008)Pascucci, Apai, Hardegree‐Ullman, Kim,
  Meyer, \& Bouwman}]{Pascucci2008}
Pascucci, I., Apai, D., Hardegree‐Ullman, E.~E., Kim, J.~S., Meyer, M.~R., \&
  Bouwman, J. 2008, The Astrophysical Journal, 673, 477

\bibitem[{Pecaut {et~al.}(2012)Pecaut, Mamajek, \& Bubar}]{Pecaut2012}
Pecaut, M.~J., Mamajek, E.~E., \& Bubar, E.~J. 2012, The Astrophysical Journal,
  746, 154

\bibitem[{Pickles(1998)}]{Pickles1998}
Pickles, A.~J. 1998, Publications of the Astronomical Society of the Pacific,
  110, 863

\bibitem[{Preibisch {et~al.}(2002)Preibisch, Brown, Bridges, Guenther, \&
  Zinnecker}]{Preibisch2002}
Preibisch, T., Brown, A. G.~A., Bridges, T., Guenther, E., \& Zinnecker, H.
  2002, The Astronomical Journal, 124, 404

\bibitem[{Raghavan {et~al.}(2010)Raghavan, McAlister, Henry, Latham, Marcy,
  Mason, Gies, White, \& ten Brummelaar}]{Raghavan2010}
Raghavan, D., {et~al.} 2010, The Astrophysical Journal Supplement Series, 190,
  1

\bibitem[{Reggiani \& Meyer(2011)}]{Reggiani2011}
Reggiani, M.~M., \& Meyer, M.~R. 2011, The Astrophysical Journal, 738, 60

\bibitem[{Riaz {et~al.}(2009)Riaz, Lodieu, \& Gizis}]{Riaz2009}
Riaz, B., Lodieu, N., \& Gizis, J.~E. 2009, The Astrophysical Journal, 705,
  1173

\bibitem[{Riaz {et~al.}(2012)Riaz, Lodieu, Goodwin, Stamatellos, \&
  Thompson}]{Riaz2012}
Riaz, B., Lodieu, N., Goodwin, S., Stamatellos, D., \& Thompson, M. 2012,
  Monthly Notices of the Royal Astronomical Society, 420, 2497

\bibitem[{Rodigas {et~al.}(2012)Rodigas, Hinz, Leisenring, Vaitheeswaran,
  Skemer, Skrutskie, Su, Bailey, Schneider, Close, Mannucci, Esposito,
  Arcidiacono, Pinna, Argomedo, Agapito, Apai, Bono, Boutsia, Briguglio, Brusa,
  Busoni, Cresci, Currie, Desidera, Eisner, Falomo, Fini, Follette, Fontana,
  Garnavich, Gratton, Green, Guerra, Hill, Hoffmann, Jones, Krejny, Kulesa,
  Males, Masciadri, Mesa, McCarthy, Meyer, Miller, Nelson, Puglisi,
  Quiros-Pacheco, Riccardi, Sani, Stefanini, Testa, Wilson, Woodward, \&
  Xompero}]{Rodigas2012}
Rodigas, T.~J., {et~al.} 2012, The Astrophysical Journal, 752, 57

\bibitem[{Scholz {et~al.}(2007)Scholz, Jayawardhana, \& Wood}]{Scholz2007}
Scholz, A., Jayawardhana, R., \& Wood, K. 2007, The Astrophysical Journal, 660,
  1517

\bibitem[{Shu {et~al.}(1994)Shu, Najita, Ostriker, Wilkin, Ruden, \&
  Lizano}]{Shu1994}
Shu, F., Najita, J., Ostriker, E., Wilkin, F., Ruden, S., \& Lizano, S. 1994,
  The Astrophysical Journal, 429, 781

\bibitem[{Sivanandam(2006)}]{Sivanandam2006}
Sivanandam, S. 2006, Proceedings of SPIE, 6269, 62690U

\bibitem[{Skemer {et~al.}(2011)Skemer, Close, Szűcs, Apai, Pascucci, \&
  Biller}]{Skemer2011}
Skemer, A.~J., Close, L.~M., Szűcs, L., Apai, D., Pascucci, I., \& Biller,
  B.~A. 2011, The Astrophysical Journal, 732, 107

\bibitem[{Skemer {et~al.}(2012)Skemer, Hinz, Esposito, Burrows3, Leisenring,
  Skrutskie, Desidera, Mesa, Arcidiacono, Mannucci, Rodigas, Close, McCarthy,
  Kulesa, Agapito, Apai, Argomedo, Bailey, Boutsia, Briguglio, Brusa, Busoni,
  Claudi, Eisner, Fini, Follette, Garnavich, Gratton, Guerra, Hill, Hoffmann,
  Jones, Krejny, Males, Masciadri, Meyer, Miller, Morzinski, Nelson, Pinna,
  Puglisi, Quanz, Quiros-Pacheco, Riccardi, Stefanini, Vaitheeswaran, Wilson,
  \& Xompero}]{Skemer2012}
Skemer, A.~J., {et~al.} 2012, The Astrophysical Journal, 753, 14

\bibitem[{Skrutskie {et~al.}(2010)Skrutskie, Jones, Hinz, Garnavich, Wilson,
  Nelson, \& Solheid}]{Skrutskie2010}
Skrutskie, M.~F., Jones, T., Hinz, P., Garnavich, P., Wilson, J., Nelson, M.,
  \& Solheid, E. 2010, Proceedings of SPIE, 7735, 77353H

\bibitem[{Skrutskie {et~al.}(2006)Skrutskie, Cutri, Stiening, Weinberg,
  Schneider, Carpenter, Beichman, Capps, Chester, Elias, Huchra, Liebert,
  Lonsdale, Monet, Price, Seitzer, Jarrett, Kirkpatrick, Gizis, Howard, Evans,
  Fowler, Fullmer, Hurt, Light, Kopan, Marsh, McCallon, Tam, {Van Dyk}, \&
  Wheelock}]{Skrutskie2006}
Skrutskie, M.~F., {et~al.} 2006, The Astronomical Journal, 131, 1163

\bibitem[{Song {et~al.}(2012)Song, Zuckerman, \& Bessell}]{Song2012}
Song, I., Zuckerman, B., \& Bessell, M.~S. 2012, The Astronomical Journal, 144,
  8

\bibitem[{Spiegel {et~al.}(2011)Spiegel, Burrows, \& Milsom}]{Spiegel2011a}
Spiegel, D.~S., Burrows, A., \& Milsom, J.~a. 2011, The Astrophysical Journal,
  727, 57

\bibitem[{Tuthill {et~al.}(2001)Tuthill, Monnier, \& Danchi}]{Tuthill2001}
Tuthill, P.~G., Monnier, J.~D., \& Danchi, W.~C. 2001, Nature, 409, 1012

\bibitem[{Urban {et~al.}(2012)Urban, Rieke, Su, \& Trilling}]{Urban2012}
Urban, L.~E., Rieke, G., Su, K., \& Trilling, D.~E. 2012, The Astrophysical
  Journal, 750, 98

\bibitem[{van Leeuwen(2007)}]{VanLeeuwen2007}
van Leeuwen, F. 2007, Astronomy \& Astrophysics, 474, 653

\bibitem[{Veras {et~al.}(2009)Veras, Crepp, \& Ford}]{Veras2009}
Veras, D., Crepp, J.~R., \& Ford, E.~B. 2009, The Astrophysical Journal, 696,
  1600

\bibitem[{Wainscoat \& Cowie(1992)}]{Wainscoat1992}
Wainscoat, R.~J., \& Cowie, L.~L. 1992, The Astronomical Journal, 103, 332

\bibitem[{White \& Ghez(2001)}]{White2001}
White, R.~J., \& Ghez, A.~M. 2001, The Astrophysical Journal, 556, 265

\bibitem[{Wright {et~al.}(2010)Wright, Eisenhardt, Mainzer, Ressler, Cutri,
  Jarrett, Kirkpatrick, Padgett, McMillan, Skrutskie, Stanford, Cohen, Walker,
  Mather, Leisawitz, Gautier, McLean, Benford, Lonsdale, Blain, Mendez, Irace,
  Duval, Liu, Royer, Heinrichsen, Howard, Shannon, Kendall, Walsh, Larsen,
  Cardon, Schick, Schwalm, Abid, Fabinsky, Naes, \& Tsai}]{Wright2010}
Wright, E.~L., {et~al.} 2010, The Astronomical Journal, 140, 1868

\end{thebibliography}

\end{document}